\documentstyle[11pt,aaspp4,psfig]{article}

\def\be{\begin{equation}}
\def\ee{\end{equation}}
\def\ba{\begin{array}}
\def\ea{\end{array}}
\def\Mesz{M\'esz\'aros}
\def\siml{\lower4pt \hbox{$\buildrel < \over \sim$}}
\def\simg{\lower4pt \hbox{$\buildrel > \over \sim$}}

\begin{document}

\title{High Energy Spectral Components in Gamma-Ray Burst Afterglows}

\author{Bing Zhang \& Peter M\'esz\'aros}
\affil{Astronomy \& Astrophysics Dept.,
 Pennsylvania State University, University Park, PA 16803}

\begin{abstract} We investigate two high energy radiation mechanisms,
the proton synchrotron and the electron inverse Compton emission, and
explore their possible signatures in the broad-band spectra and in the
keV to GeV light curves of gamma-ray burst afterglows. We develop a
simple analytical approach, allowing also for the effects of
photon-photon pair production, and explore the conditions under which
one or the other of these components dominates.
We identify three parameter space regions where different spectral 
components dominate: (I) a region where the proton synchrotron and other
hadron-related emission components dominate, which is small; (II) a
region where the electron inverse Compton component dominates, which is
substantial; (III) a third substantial region where electron  synchrotron
dominates. We discuss the prospects and
astrophysical implications of directly detecting the inverse Compton
and the proton high energy components in various bands, in particular
in the GeV band with future missions such as {\em GLAST}, and in the
X-ray band with {\em Chandra}. We find that regime II parameter space
is the most favorable regime for high energy emission. The inverse
Compton component is detectable by {\em GLAST} within hours for bursts
at typical cosmological distances, and by {\em Chandra} in days if
the ambient density is high.
\end{abstract}


\section{Introduction}
\label{sec:intro}

Gamma-ray Burst (GRB) afterglows have been detected mainly at longer
wavebands, from X-rays to radio, extending up to months after the
burst triggers (e.g. van Paradijs, Kouveliotou \& Wijers 2000).  These
long-lived afterglow emissions are generally interpreted within the
fireball shock model by the synchrotron emission of external
shock-accelerated relativistic electrons.  The broadband electron
synchrotron spectrum (Sari, Piran \& Narayan 1998) has proven to be a
useful paradigm to study afterglow lightcurves in the X-ray, optical,
and radio band, and to constrain various unknown parameters.  On the
other hand, there are other high energy emission components, whose
role in determining the long wavelength afterglows may be secondary or
of limited duration, but which may, for some time, dominate the high
energy, X-ray to GeV spectrum of the afterglows. These include the
synchrotron self-inverse Compton emission (IC) of the electrons
(\Mesz, Rees \& Papathanassiou 1994; Waxman 1997; Panaitescu \& \Mesz~
1998; Wei \& Lu 1998, 2000; Dermer, B\"ottcher \& Chiang 2000a;
Dermer, Chiang \& Mitman 2000b; Panaitescu \& Kumar 2000; Sari \& Esin
2001), the proton synchrotron emission (Vietri 1997; B\"ottcher \&
Dermer 1998; Totani 1998) as well as some other hadron-related
emission components (B\"ottcher \& Dermer 1998). The emission from
such high energy mechanisms in the GeV band may have been detected by
{\em EGRET} in GRB 940217 (Hurley et al. 1994) and probably also in
some other GRBs, and will be detectable by next generation GeV
$\gamma$-ray missions such as {\em GLAST}.  In the X-ray band, the
electron IC component has been pointed out as potentially detectable
at a later time (Sari \& Esin 2001).

Previous afterglow snapshot spectral fits to multi-wavelength data up
to X-rays using the synchrotron emission component (e.g. Wijers \&
Galama, 1999; Panaitescu \& Kumar 2000; van
Paradijs, Kouveliotou \& Wijers, 2000; Freedman \& Waxman 2001) have
constrained some of the fireball shock parameters (e.g. the electron
and magnetic field energy equipartition parameters $\epsilon_e$,
$\epsilon_B$), which are found to vary over a wide range. Because of
the relative looseness of these fits, there is a need to tighten the
model constraints. A promising way to do this is by extending upward
the frequency range over which the snapshot spectra are fitted, and to
investigate the relative importance and additional constraints imposed
by the high energy radiation components, over a wider region of
parameter phase space, which is the subject of this paper.  For this
purpose, we use a simple analytic approach to describe the various
spectral components. Following a brief general treatment of the
particle distribution and the synchrotron spectrum of the electrons
(\S
\ref{sec:cool}), we discuss the proton synchrotron spectral component
(\S \ref{sec:sync}) and the electron inverse Compton component (\S
\ref{sec:ic}), and compare their importance relative to the electron
synchrotron emission in the frequency regime below the electron
synchrotron cut-off. Above this cut-off, the relative importance
between the electron's IC and the hadron-related components is also
discussed. In \S \ref{sec:gammagamma} we incorporate the high energy
absorption due to $\gamma-\gamma$ pair production in an analytic
manner. We then present several examples of the broad band spectra of
GRB afterglows within different parameter regimes in \S
\ref{sec:results}, and discuss the detectability of these high energy
spectral components within various bands, especially in the X-ray band
and the GeV band. Finally we summarize our findings in \S
\ref{sec:sum}.

\section{Proton and electron distributions and cooling}
\label{sec:cool}

We assume that for negligible radiative losses both electrons and
protons are shock accelerated to a single relativistic power law
distribution of index $p$. Denoting the particle species by the
subindex ``x'' (which indicates ``e'' for electrons and ``p'' for
protons), after a certain time the particle distribution becomes a
broken power law depending on the relative ordering between the
injection minimum energy, $\gamma_{m,x}$, and the energy at which the
particles cool radiatively in an expansion timescale, $\gamma_{c,x}$.
Given a maximum particle energy $\gamma_{u,x}$ at which the
acceleration time equals the energy loss time, the particle
distribution is $N(\gamma_x)\propto \gamma_x^{-p}$ for
$\gamma_{m,x}<\gamma_x<\gamma_{c,x}$, and $N (\gamma_x)\propto
\gamma_{x}^{-p-1}$ for $\gamma_{c,x}<\gamma_x<\gamma_{u,x}$, if
$\gamma_{m,x}<\gamma_{c,x}$ (slow cooling regime); and is
$N(\gamma_x)\propto \gamma_x^{-2}$ for
$\gamma_{c,x}<\gamma_x<\gamma_{m,x}$, and $N (\gamma_x)\propto
\gamma_{x}^{-p-1}$ for $\gamma_{m,x}<\gamma_x<\gamma_{u,x}$, if
$\gamma_{m,x}>\gamma_{c,x}$ (fast cooling regime).  For $p>2$, the
mean particle energy is $\bar\gamma_x=[(p-1)/(p-2)]\gamma_{m,x}$.
Hereafter we assume for simplicity the same power law index $p$ for
both electrons and protons. Assuming that $\zeta_p n$ and $\zeta_e n$
are the number densities of the shock-accelerated power-law protons
and electrons, respectively, where $\zeta_p$ and $\zeta_e$ are the
injection fractions and $n$ is the number density of the pre-shock
thermal medium, one can define the energy portion contained in the
power-law distributed electrons and protons with respect to the total
energy behind the shock to be $\epsilon_e=\zeta_e (m_e/m_p)
(\bar\gamma_e /\Gamma)$, and $\epsilon_p =\zeta_p(\bar\gamma_p
/\Gamma)$, respectively, where $\Gamma$ is the bulk Lorentz factor of
the blast wave. Consequently, the minimum energy for the power-law
distributed electrons and the protons are
\footnote{We note that the larger minimum proton energy estimates
$\gamma_{m,p}=\Gamma$ adopted, e.g., by Vietri 1997; B\"ottcher \&
Dermer 1998 and Totani 1998, when $p$ is close to 2 as assumed in
those references, lead to larger proton synchrotron flux levels than
those obtained using equation (\ref{gammam}), because of the higher
${\bar
\gamma_p}$ involved.}
\be
\gamma_{m,e}=\left({\epsilon_e\over \zeta_e}\right)\left({m_p \over m_e}
\right) \left({p-2 \over p-1}\right) \Gamma, ~~~ \gamma_{m,p} =
\left({\epsilon_p \over \zeta_p}\right)\left({p-2 \over p-1}\right)
\Gamma .
\label{gammam}
\ee
The above equations are for $p>2$, which for simplicity will be
assumed to be valid. For $p$ getting closer to 2, the $(p-2)/(p-1)$
factor is no longer precise, and when $p=2$, it is replaced by ${\rm
ln}^{-1}(\gamma_{u,x} /\gamma_{m,x})$. However, the imprecise of
eq.(\ref{gammam}) at $p=2$ does not influence the precise of the
discussions in \S\ref{sec:sync} (e.g. eq.[{\ref{num}]) since both
correction factors for the proton and the electron components are
cancelled out.  The cooling energy $\gamma_{c,x}$ is defined by
equating the comoving adiabatic expansion time $t'_{ad} \sim r/(\Gamma
c) \sim \Gamma t$ (where $\Gamma$ is the bulk Lorentz factor of the
blastwave at the radius $r$, $t \sim \Gamma^{-2}(r/c)$ is the
expansion time in the observer frame) to the comoving radiation
cooling time
$t'_c=[(t'_{sy})^{-1}+(t'_{IC})^{-1}]^{-1}=t'_{sy}/(1+Y)$, where
$t'_{sy}$ and $t'_{IC}$ are the synchrotron and the IC cooling time,
and $Y=t'_{sy}/t'_{IC}=(-1+\sqrt{1+4\eta \epsilon_x/\epsilon_B})/2$ is
the Compton factor (e.g. Panaitescu \& Kumar 2000; Sari \& Esin 2001).
Here redshift corrections have been ignored for simplicity, $\eta={\rm
Min}[1, (\gamma_{m,x}/\gamma_{c,x})^{p-2}]$ is the fraction of the
particle energy that is radiated away via both synchrotron and inverse
Compton (Sari \& Esin 2001), and $\epsilon_B$ is the fraction of the
magnetic energy density with respect to the total energy density
behind the shock, so that the comoving magnetic field $B'=\Gamma
c(32\pi nm_p\epsilon_B)^ {1/2}$. For $\eta\epsilon_x/\epsilon_B \ll
1$, one has $Y\sim \eta\epsilon_x/\epsilon_B\ll1$ and $t'_c \sim
t'_{sy}$, and the IC cooling is not important. Alternatively, when
$\eta\epsilon_x/\epsilon_B \gg 1$, IC cooling is important, and $t'_c
\sim (\eta\epsilon_e/\epsilon_B)^{-1/2} t'_{sy}$ (Sari \& Esin 2001).
The critical energy above which the particle species $x$ cool in an
expansion time is $\gamma_{c,x}=(1+Y_x)^{-1}(6\pi m_x
c/\Gamma\sigma_{T,x} {B'}^2 t)$, where $t'_{sy}=(\gamma_x m_x
c^2)/[(4/3)\sigma_{T,x} c\gamma_x^2 (B'^2/8\pi)]$ is used,
$\sigma_{T,x}=(8\pi/3)(e^2/m_xc^2)^2$ is the Thomson cross section for
particles of mass $m_x$, and $\sigma_{T,p}/\sigma_{T,e}=(m_e/m_p)^2$.
A caveat is that, by adopting 
$Y=(-1+\sqrt{1+4\eta \epsilon_x/\epsilon_B})/2$, one has implicitly 
assumed $L_{IC} / L_{sy} \sim U_{sy}/U_B$, where $U_{sy}$, $U_B$ are 
the synchrotron radiation and magnetic field energy densities,
respectively. Strictly speaking, this is valid only when the IC
cooling occurs in the classical Thomson regime. In practice, high energy
electrons may cool in the Klein-Nishina regime, especially at the
early afterglow phase. The cooling frequency in the above analytic
treatment is then a crude approximation if the energy where the
electron cool is in the K-N regime, and more careful treatment ought
to be made. As discussed more in \S \ref{sec:ic}, the analytic
treatment presented in this paper is adequate at one hour or
even earlier after the burst trigger, but may be crude for the
prompt phase.

The maximum energy $\gamma_{u,x}$ is defined by equating the comoving
acceleration time $t'_{\rm acc} \equiv 2\pi\alpha r_{\rm L}/c
=2\pi\alpha\gamma_{u,x} m_xc/eB'$ (where $r_{\rm L}$ is the Larmor
radius, and $\alpha$ is a factor of the order of unity for
relativistic shocks) to the minimum of the comoving adiabatic cooling
time $t'_{ad}$ and the comoving radiation cooling time $t'_c$. This
gives $\gamma_{u,x}= {\rm Min}\{(2\pi\alpha)^{-1}(eB'/m_xc)(\Gamma t),$
$[3e/\alpha B'
\sigma_{T,x} (1+Y_x)]^{1/2} \}$, where for protons the first part of
the equation applies, while for the electrons it is the latter.

The electron synchrotron emission spectrum is a broken power law
separated by three characteristic frequencies: the self-absorption
frequency $\nu_{a,e}$, the characteristic frequency for the minimum
energy particles, $\nu_{m,e} \simeq ({4\over3}\Gamma){3\over
4\pi}{eB'\over m_ec} \gamma_{m,e}^2$, and the cooling frequency,
$\nu_{c,e} \simeq ({4\over3}\Gamma){3\over 4\pi}{eB'\over m_ec}
\gamma_{c,e}^2$. Similar expressions apply for protons, except that
synchrotron absorption and cooling are less important than in
electrons.  For both $p$ and $e$ the maximum particle energy
$\gamma_{u,x}$ defines a synchrotron cut-off frequency $\nu_{u,x}
\simeq ({4\over3}\Gamma) {3\over 4\pi}{eB'\over m_x c}
\gamma_{u,x}^2$. In the above expressions, the factor (4/3) gives a
more precise conversion from the comoving frame to the observer frame
(Wijers \& Galama 1999).  The spectral indices of the four segments
ordering from low to high in frequency are [2, 1/3, -1/2, -p/2] for
the fast-cooling ($\nu_{c,e} < \nu_{m,e}$) regime, or [2, 1/3,
-(p-1)/2, -p/2] for the slow-cooling ($\nu_{c,e} > \nu_{m,e}$) regime,
respectively (Sari et al. 1998).

\section{Synchrotron components}
\label{sec:sync}

A potentially interesting high energy emission component of GRB
afterglows is the synchrotron emission from the shock-accelerated
relativistic protons (Vietri 1997; B\"ottcher \& Dermer 1998; Totani
1998, 2000). Compared to electrons, the protons are inefficient
emitters due to their much larger mass. The ratio of the
characteristic synchrotron frequencies of protons and electrons is
\be
{\nu_{m,p} \over \nu_{m,e}} = \left({\gamma_{m,p}\over \gamma_{m,e}}
\right)^2 \left({m_e\over m_p}\right)
=\left({\epsilon_p/\zeta_p \over \epsilon_e/\zeta_e} \right)^2
\left({m_e\over m_p}\right)^3,
\label{num}
\ee
where eq.(\ref{gammam}) is used.  The peak synchrotron flux for the
emission of particle x is $F_{\nu,max,x} \propto n_x P'_{\nu'_{m,x}}
\propto n_x/m_x$, where $P'_{\nu'_{m,x}} =\phi_o(\sqrt{3}e^3B'/m_x
c^2)$, and $\phi_o\sim 0.6$ (Wijers
\& Galama, 1999). This is independent of whether the peak is at
$\nu_{m,x}$ or at $\nu_{c,x}$. Thus the ratio of the peak flux of the
two particle species is
\be
{F_{\nu,max,p} \over F_{\nu,max,e}} =
\left({\zeta_p\over \zeta_e}\right) \left({m_e\over m_p}\right).
\label{Fnum}
\ee
We see that both $\nu_{m,p}$ and $F_{\nu,max,p}$ are much smaller than
$\nu_{m,e}$ and $F_{\nu,max,e}$, respectively, indicating that the
proton emission component is usually buried under the electron
emission component. However, the ratio of the cooling frequencies is
\be
{\nu_{c,p}\over\nu_{c,e}}=\left(1+Y_e \over 1+Y_p\right)^2
\left({m_p\over m_e}\right)^6
\label{nuc}
\ee
(see \S\ref{sec:cool}), which means that protons barely cool while
electrons cool rapidly. This allows, in some cases, the proton
component to dominate over the electron component at high frequencies.

We consider now a certain observation band $\Delta \nu$ around $\nu$
which is above both the cooling frequency and the characteristic
synchrotron frequency of the electrons, but is below the electrons'
synchrotron cut-off frequency, i.e. $\nu_{c,e} ({\rm and}~
\nu_{m,e}) < \nu <\nu_{u,e}$. Regardless of the relative ordering of
$\nu_{m,e}$ and $\nu_{c,e}$ (slow-cooling or fast cooling), the
electron synchrotron flux at this frequency is $F_{\nu,e} (\nu) =
(\nu/\nu_{m,e})^{-(p-1)/2} F_{\nu,max,e}\cdot(\nu_{c,e}/\nu)^{1/2}$.
If $\nu < \nu_{u,p}$ is also satisfied, the flux of the proton
synchrotron component at the same frequency is $F_{\nu,p} (\nu)
=({\nu/\nu_{m,p}})^{-(p-1)/2}F_{\nu,max,p}$, since
$\nu_{m,p}<\nu_{m,e}$ and usually $\nu_{c,p} \simg \nu_{u,p}$.  Thus
\be
{F_{\nu,p} (\nu) \over F_{\nu,e} (\nu)} = \left({F_{\nu,max,p} \over
F_{\nu,max,e}} \right)\left({\nu_{m,p} \over \nu_{m,e}}\right)^{(p-1)/
2} \left({\nu\over\nu_{c,e}} \right)^{1/2} =\left({\epsilon_p\over
\epsilon_e}\right)^{(p-1)}
\left({\zeta_e\over \zeta_p}\right)^{(p-2)}
\left({m_e \over m_p}\right)^{(3p-1)/2}
\left({\nu \over \nu_{c,e}}\right)^{1/2}.
\label{Fpe}
\ee
We see that $\nu$ must be $\gg \nu_{c,e}$ to make the proton component
show up in the spectrum, and that a large $\epsilon_p$ and a small
$\epsilon_e$ favor the proton component. The dependence on the
injection parameters $\zeta_e$ and $\zeta_p$ is weak for $p$ close to
2.

For the purposes of numerical examples we use below a standard
scenario (e.g. \Mesz~ \& Rees 1997) in which a blast wave with total
energy per solid angle ${\cal E}=E/\Omega= 10^{52} {\rm ergs/Sr}~
{\cal E}_{52}$ expands into a uniform interstellar medium of particle
number density $n~{\rm cm^{-3}}$, and $\Omega$ is a putative jet
opening solid angle. (In follows we will not discuss the jet
dynamics. This does not influence our discussions on the early GeV
afterglows, but may quantitatively, although not qualitatively, change
the discussions of the late X-ray afterglows, if the transitions occur
after the jet breaks. A wind-like external medium with $n\propto
r^{-2}$ may be also incorporated in a similar way but we do not
discuss this here). For the uniform external medium case, the blast
wave starts to decelerate at a radius $r_{dec} \sim (3{\cal E}/4\pi
nm_p c^2\Gamma_0^2)^{1/3}\sim 2.6\times 10^{16} {\rm cm}~ {\cal
E}_{52}^{1/3} n^{-1/3}\Gamma_{0,300}^{-2/3}$, at an observer time
$t_{dec}\sim (r_{dec}/4\Gamma_0^2 c) \cdot (1+z) \sim 2.4 {\rm
s}~{\cal E}_{52}^{1/3} n^{-1/3}\Gamma_{0,300}^{-8/3}(1+z)$ s. Here
$\Gamma_0=300\Gamma_{0,300}=E/M_0c^2$ is the initial bulk Lorentz
factor of the shocked materials at $t=t_{dec}$ and the factor $(1+z)$
reflects the cosmological time dilation effect.  After collecting
enough interstellar medium materials, the blast wave will be
decelerated self-similarly
and its advance is measured by $r(t)\sim [12{\cal E}c t/(1+z)/ 4\pi
nm_pc^2]^{1/4} =1.6\times 10^{17}{\rm cm~}({\cal
E}_{52}/n)^{1/4}[t_h/(1+z)]^{1/4}$, while the bulk Lorentz factor
decays as $\Gamma(t) \sim[3{\cal E}(1+z)^3/4^3\cdot 4\pi n m_pc^5t^3]
^{1/8}=19.4({\cal E}_{52}/n)^{1/8}[t_h/(1+z)]^{-3/8}$.  Here $r=
4\Gamma^2 ct (1+z)$ has been adopted, and $t_h$ is the earth observer
time in unit of hours. The comoving magnetic field is $B'= 7.5 {\rm
G~}
\epsilon_{B}^{1/2} {\cal E}_{52}^{1/8}n^{3/8}[t_h/(1+z)]^{-3/8}$ G.
The breaks in the electron energy distribution are then given by
\begin{eqnarray}
  \gamma_{m,e}= & 5.9\times 10^3 (\epsilon_{e} / \zeta_e) ({\cal
  E}_{52}/n)^{1/8} [t_h/(1+z)]^{-3/8},
\label{gamme}\\
\gamma_{c,e} = & 1.96 \times 10^2 (1+Y_e)^{-1}
\epsilon_{B}^{-1} {\cal E}_{52}^{-3/8} n^{-5/8} [t_h/(1+z)]^{1/8}.
\label{gamce}
\end{eqnarray}
In eq.(\ref{gamme}) and hereafter $p=2.2$ for both electrons and
protons has been adopted to calculate numerically the coefficients.
The $p$-dependences in the power indices, if necessary, will be still
retained.  The observed characteristic electron synchrotron frequency
is then $\nu_{m,e}=2.9\times 10^{16}{\rm Hz} (\epsilon_e/\zeta_e)^2
\epsilon_{B}^{1/2}{\cal E}_{52}^{1/2} t_h^{-3/2} (1+z)^{1/2}$, while
that for the protons is $\nu_{m,p} =4.6\times 10^6 {\rm
Hz}(\epsilon_p/\zeta_p)^2\epsilon_{B}^{1/2} {\cal E}_{52}^{1/2}
t_h^{-3/2} (1+z)^{1/2}$.  The electron cooling frequency is
\be
\nu_{c,e} =3.1\times 10^{13}{\rm Hz} (1+Y_e)^{-2}\epsilon_{B}^{-3/2}
{\cal E}_{52}^{-1/2}n^{-1} t_h^{-1/2} (1+z)^{-1/2},
\label{nuce}
\ee
and the cut-off maximum frequencies for the electrons and the protons
are
\begin{eqnarray}
\nu_{u,e} =& 2.3\times
10^{23} {\rm Hz}~ \alpha^{-1}(1+Y_e)^{-1}({\cal E}_{52}/n)^{1/8}
t_h^{-3/8}(1+z)^{-5/8}, \label{nuue}\\
\nu_{u,p}= &2.8\times 10^{23}
{\rm Hz}~ \alpha^{-2} \epsilon_B^{3/2} ({\cal E}_{52} n)^{3/4}
t_h^{-1/4}(1+z)^{-3/4}.
\label{nuup}
\end{eqnarray}
Finally, the maximum electron synchrotron flux is $F_{\nu,max,e} =
[(4\pi/3)r^3 \zeta_e n \Gamma P'_{\nu'_{m,e}} /4\pi D^2] (1+z) =
(\phi_o \sqrt{3}/3) (e^3\zeta_e n/mc^2 D^2) \Gamma B' r^3 (1+z) = 29
{\rm mJy} \zeta_e\epsilon_{B}^{1/2} n^{1/2}{\cal E}_{52} D_{28}^{-2}
(1+z)$, where $D \equiv D(z) =10^{28}{\rm cm}D_{28}$ is the proper
distance of the source (which, depending on the cosmological model, is
also redshift dependent). Similarly, the maximum proton synchrotron
flux is $F_{\nu,max,p}=15.7{\rm \mu
Jy}\zeta_p\epsilon_{B}^{1/2}n^{1/2} {\cal E}_{52} D_{28}^{-2} (1+z)$.

For $\nu<\nu_{u,e}$, the condition for the proton synchrotron
component to overcome the electron component is $F_{\nu,p}(\nu_{u,p})
> F_{\nu,e}(\nu_{u,p})$. Using equations (\ref{Fpe}), (\ref{nuce}) and
(\ref{nuup}), this translates into
\be
 (1+Y_e)^{2/3}\epsilon_B > 594 (\epsilon_e/\epsilon_p)^{2(p-1)/3}
 (\zeta_p/\zeta_e)^{2(p-2)/3} \alpha^{2/3}{\cal E}_{52}^{-5/12}
 n^{-7/12} [t_h/(1+z)]^{-1/12},
\label{1}
\ee
which is shown as line 1 in Fig. 1.

Another condition for the competition between the $p$ and $e$
components which might be thought to be relevant is
$\nu_{u,p}>\nu_{u,e}$, which corresponds to $ (1+Y_e)^{2/3}\epsilon_B
> 0.88 \alpha^{2/3}{\cal E}_{52}^{-5/12}$ $n^{-7/12}
[t_h/(1+z)]^{-1/12}$. However, this condition may not be essential,
since as shown by B\"ottcher \& Dermer (1998) there can be other
proton-induced electromagnetic signals extending above this energy, of
level comparable to or lower than the proton synchrotron
component. These include the synchrotron radiation from the positrons
produced by $\pi^{+}$ decay and the $\gamma$-rays produced directly
from $\pi^{0}$ decay. These components may be regarded as an extension
to the proton synchrotron component, which would stick out above the
electron synchrotron component even if $\nu_{u,p}<\nu_{u,e}$. However,
these proton components will compete with the electron IC component,
which we discuss in \S \ref{sec:ic}.

\section{Inverse Compton component}
\label{sec:ic}

The inverse Compton (IC) is mainly important for the electron
component, both for the electron cooling and for forming a separate
high energy emission component\footnote{The IC of the protons is not
important since $\eta_{p}=(\gamma_{m,p}/\gamma_{c,p})^{p-2} \ll 1$.}.
The condition for the IC cooling to be important is
$\eta\epsilon_e/\epsilon_B>1$, which has been explicitly addressed by
Sari \& Esin (2001). Here we investigate the condition that the IC
emission component overtakes the electron and/or the proton
synchrotron component.

The IC component has a similar spectral shape to the synchrotron
component, but its low energy peak is Lorentz-boosted by roughly a
factor of $\gamma_{m,e}^2$, while the frequency spread where it is
important is stretched out, by comparison to the spread of the
synchrotron spectrum, extending between the boosted characteristic and
cooling frequencies.  Sari \& Esin (2001) have explicitly presented
analytic expressions for the IC spectral component, and found that the
power law approximation is no longer accurate at $\nu > \gamma_m^2
\nu_{m,e}$, since electrons with a range of Lorentz factors between
$\gamma_m$ and $\gamma_c$ contribute equally to the emission at each
frequency. For the convenience of the following discussions, we will
still adopt the broken power law approximation to perform
order-of-magnitude estimations, bearing in mind that more accurate
expressions would be necessary in more detailed calculations.

In this approximation, the IC spectral component can be represented
(Sari \& Esin, 2001) by a four-segment broken power law with power
indices ordered from low to high frequency of [1, 1/3, -(p-1)/2, -p/2]
in the slow-cooling regime, or of [1, 1/3, -1/2, -p/2] in the
fast-cooling regime. The break frequencies are $\nu_{a,e}^{\rm
IC}=\gamma_{m,e}^2\nu_{a,e}$, $\nu_{m,e}^{\rm IC} \simeq
\gamma_{m,e}^2\nu_{m,e}$, and the IC cooling frequency $\nu_{c,e}^{\rm
IC}\simeq \gamma_{c,e}^2\nu_{c,e}$.  The maximum flux of the inverse
Compton component is roughly a factor
$(u'_{ph}/u'_B)(\nu_{m,e}/\nu^{\rm IC}_{m,e})$ of that of the
synchrotron component, where $u'_{ph}\simeq (4/3)c\sigma_{T,e} u'_B
\gamma_e^2
\cdot (r/\Gamma c) \cdot (4\Gamma \zeta_e n)$ and $u'_B = B'^2/8\pi$ are the
comoving synchrotron photon and magnetic field energy densities,
respectively.  This gives
\be
{F_{\nu,max,e}^{\rm IC}\over F_{\nu,max,e}}\sim {16\over 3}
\sigma_{T,e}
\zeta_e nr=3.5\times 10^{-7}\zeta_e n r_{17}=5.7\times 10^{-7}\zeta_e
{\cal E}_{52}^{1/4} n^{3/4} [t_h/(1+z)]^{1/4}~.
\label{FmIC}
\ee
This shows that generally the IC component can only overtake the
synchrotron component beyond the synchrotron component's cooling
break, but before the IC component's cooling break. In the
slow-cooling regime, for a frequency $\nu$ satisfying $\nu_{c,e}\leq
\nu \leq
\nu_{u,e}$ and $\nu_{m,e}^{\rm IC} \leq \nu \leq \nu_{c,e}^{\rm IC}$,
the flux ratio of the IC and the synchrotron components is
\be
{F^{\rm IC}_{\nu,e} (\nu) \over F_{\nu,e} (\nu)} =
\left({F^{\rm IC}_{\nu,max,e}\over F_{\nu,max,e}}\right)
\left({\nu^{\rm IC}_{m,e} \over \nu_{m,e}}\right)^{(p-1)/2}
\left({\nu \over \nu_{c,e}}\right)^{1/2}.
\label{FIC1}
\ee
Alternatively, in the fast-cooling regime, for a frequency $\nu$
satisfying $\nu_{m,e}\leq \nu \leq
\nu_{u,e}$ and $\nu_{c,e}^{\rm IC} \leq \nu \leq \nu_{m,e}^{\rm IC}$,
the flux ratio of the IC and the synchrotron components is
\be
{F^{\rm IC}_{\nu,e} (\nu) \over F_{\nu,e} (\nu)} =
\left({F^{\rm IC}_{\nu,max,e}\over F_{\nu,max,e}}\right)
\left({\nu^{\rm IC}_{c,e} \over \nu_{c,e}}\right)^{1/2}
\left({\nu \over \nu_{m,e}}\right)^{(p-1)/2}.
\label{FIC2}
\ee
Similarly to the proton synchrotron case, the conditions that the IC
component overcomes the synchrotron component is $F_{\nu,e}^{\rm
IC}(\nu_{c,e}^{\rm IC}) > F_{\nu,e}(\nu_{c,e}^{\rm IC})$ for the
slow-cooling case, or $F_{\nu,e}^{\rm IC}(\nu_{m,e}^{\rm IC}) >
F_{\nu,e}(\nu_{m,e}^{\rm IC})$ for the fast-cooling case, and both
conditions can be simplified to
\be
{16\over 3} \sigma_{T,e}\zeta_e nr \gamma_{c,e}\gamma_{m,e}^{(p-1)}>1.
\label{FIC}
\ee
Using eqs.(\ref{gamme}) and (\ref{gamce}), this IC-dominance condition
over electron synchrotron can be re-expressed as
\be
(1+Y_e)\epsilon_B < 3.8 (\epsilon_e/\zeta_e)^{(p-1)} \zeta_e ({\cal
E}_{52}/n)^{(p-2)/8} [t_h/(1+z)]^{-3(p-2)/8}.
\label{2}
\ee
This is the line labeled 2 in Fig. 1.  We see that a large
$\epsilon_e$ or a small $\epsilon_B$ makes the IC component more
prominent, since a large $\epsilon_e$ enhances $\gamma_{m,e}$ and a
small $\epsilon_B$ tends to increase $\gamma_{c,e}$ due to the
inefficient synchrotron cooling.  A denser medium also favors the IC
component (eq.[\ref{FmIC}]).  One comment is that if one takes into
account the flux increase above $\nu_{m,e}^{\rm IC}$ due to the
logarithmic term from the scattering contributions of the different
electrons (Sari \& Esin 2001), the IC dominance condition is less
stringent than (\ref{2}).

The cut-off energy in the IC component is defined by $\nu_{u,e}^{\rm
IC}= {\rm Min} (\gamma_{u,e}^2\nu_{u,e}, \nu_{_{{\rm KN},e}}^{\rm IC})$,
where $\nu_{_{{\rm KN},e}}^{\rm IC}$ is the Klein-Nishina limit. A rough
estimate of this frequency is given by $\nu_{_{{\rm KN},e}}^{\rm IC}
\sim \gamma_{_{{\rm KN},e}}^2 \nu_{_{{\rm KN},e}}$, where $\nu_{_{{\rm
KN},e}}\simeq ({4\over3}\Gamma){3\over 4\pi}{eB'\over
m_ec}\gamma_{_{{\rm KN},e}}^2$,
and $\gamma_{_{{\rm KN},e}} \sim mc^2/ {\rm Min}(h\nu'_{m,e},
h\nu'_{c,e})={\rm Max}(\gamma_{_{{\rm KN},e,1}}, \gamma_{_{{\rm
KN},e,2}})$, where $\gamma_{_{{\rm KN},e,1}}=1.1\times 10^5
(\xi_e/\epsilon_e)^2 \epsilon_B^{-1/2} {\cal E}_{52}^{-3/8} n^{-1/8}
t_h^{9/8} (1+z)^{-1/8}$, and $\gamma_{_{{\rm KN},e,2}}=1.0 \times 10^8
(1+Y_e)^2 \epsilon_B^{3/2} 
{\cal E}_{52}^{5/8} n^{7/8} t_h^{1/8} (1+z)^{7/8}$. In any case, the
condition $\nu_{u,e}^{\rm IC}\gg \nu_{u,e}$ always holds, which means 
that above $\nu_{u,e}$ the IC component always sticks out.
By comparing both  $\gamma_{_{{\rm KN},e,1}}$ and $\gamma_{_{{\rm
KN},e,2}}$ with $\gamma_{c,e}$ (eq.[\ref{gamce}]) and noticing their
temporal dependences, we can see that at $t_h=1$, both $\gamma_{_{{\rm
KN},e,1}}$ and $\gamma_{_{{\rm KN},e,2}}$ are far greater than
$\gamma_{c,e}$, and only $\gamma_{_{{\rm KN},e,1}}$ may becomes smaller
than $\gamma_{c,e}$ at the prompt phase. This means that throughout
the afterglow phase, the scatterings with the peak-flux background
photons ($\nu'_{m,e}$ for slow-cooling and $\nu'_{c,e}$ for
fast-cooling) for the cooling energy electrons are well in the
classical regime, and at $t_h \sim 1$ or even earlier, the scatterings
with the peak-energy-flux background photons ($\nu'_{c,e}$ for
slow-cooling and $\nu'_{m,e}$ for fast-cooling) for the cooling energy
electrons are also in the classical regime. This justifies the
approximate analytic cooling treatment adopted in this paper, at least
for $t_h > 1$. For earlier afterglows, especially at the prompt phase,
the cooling frequency estimate may be still viable, but the spectrum
above the cooling frequency ought to be treated more carefully by
taking into account the Klein-Nishina modifications.

In the $\epsilon_e, \epsilon_B$ space (Fig.1), we see that the regions
defined by (\ref{1}) and (\ref{2}) do not overlap. This means that for
frequencies $\nu <\nu_{u,e}$, given a certain set of parameters, only
one or the other high energy component (IC or proton synchrotron)
competes with the electron synchrotron component. We note that there
is also a substantial region of parameter space in which neither of
these high energy components can be dominant. For frequencies $\nu>
\nu_{u,e}$, on the other hand, there is an overlap of the proton
dominated phase space region with the IC dominated phase space region
in the $\epsilon_e, \epsilon_B$ plane.  We thus need to compare the
relative importance of these two components. Below, we choose the
slow-cooling case as an example. The fast-cooling case can be
discussed in a similar way and results in qualitatively similar
conclusions.

Within the slow-cooling regime, there are two cases.  If
$\nu_{u,e}<\nu <\nu_{c,e}^{\rm IC}$, the IC emission is important, and
one can use eqs. (\ref{Fpe}), (\ref{FmIC}) and (\ref{FIC1}) to define
the condition that the proton component overcomes the IC
component\footnote{Here we have
assumed a $-(p-1)/2$ slope for the proton component extending to
infinity.  When $\nu>\nu_{u,p}$, this is an over-estimate for the
other hadron-related components (B\"ottcher \& Dermer 1998). Thus the
real criterion is even more stringent than (\ref{3}).},
i.e., $F_{\nu,p}(\nu)/F_{\nu,e}^{\rm IC}(\nu)=(\epsilon_p/\epsilon_e)
^{(p-1)} (\zeta_e/\zeta_p)^{(p-2)}$ $(m_e/m_p)^{(3p-1)/2} / [(16/3)
\sigma_{T,e} \zeta_e n r \gamma_{m,e}^{(p-1)}]>1$. This can be
translated into
\be
\epsilon_e< 8.1\times
10^{-4}\epsilon_p^{1\over 2}\zeta_e^{p-2\over
p-1}\zeta_p^{-{p-2\over2(p-1)}} {\cal E}_{52}^{-{p+1\over
16(p-1)}}n^{-{7-p\over16(p-1)}} [t_h/(1+z)]^{3p-5\over16(p-1)}.
\label{3}
\ee
This is line 3 in Fig. 1, to the left of which proton synchrotron
dominates over IC, which again is a relatively small phase space
region in terms of $\epsilon_e$ (but larger than the region left of
line 1, where proton synchrotron dominates over electron synchrotron).

The second case in the slow-cooling regime is for $\nu >
\nu_{c,e}^{\rm IC}$, and here the competition between the two
components is more difficult to quantify.  The fact that usually
$\nu_{u,p} \ll \nu_{u,e}^{\rm IC}$ would seem to indicate that the
proton component is not important. However, as B\"ottcher \& Dermer
(1998) have shown, photo-meson interactions between the relativistic
protons and the low energy photon spectrum lead to additional
hadron-related spectral components at high energies, with a $\nu
F_\nu$ level which can be a substantial fraction of that of the proton
synchrotron component, but extending to much higher energies.  For our
purposes it is not necessary here to compute these components in
detail, assuming instead as a rough estimate that beyond the maximum
proton synchrotron frequency $\nu_{u,p}$ the extended hadron component
$\nu F_\nu$ flux level represents a fraction $k=0.1 k_{-1}$ of the
proton synchrotron flux level at the proton synchrotron cut-off,
$F_{\nu,p}(\nu_{u,p})= F_{\nu,max,p}
(\nu_{u,p}/\nu_{m,p})^{-(p-1)/2}$, so that one has
$F_{\nu,p}(\nu)/F_{\nu,p}(\nu_{u,p})=0.1 k_{-1} (\nu_{u,p}/\nu)$. This
is consistent with B\"ottcher \& Dermer's Fig.1, and in general $k=0.1
k_{-1}$ represents an overestimate of the flux level from these
hadron-related components. We then compare $F_{\nu,p}(\nu)$ to
$F_{\nu,e}^{\rm IC}(\nu)=F_{\nu,max,e}^{\rm IC} (\nu^{\rm
IC}_{c,e}/\nu^{\rm IC}_{m,e})^{-(p-1)/2} (\nu/\nu^{\rm
IC}_{c,e})^{-p/2}$. Again with the help of (\ref{Fpe}), (\ref{FmIC})
and (\ref{FIC1}), and noticing $F_{\nu,e}^{\rm IC}(\nu)
/F_{\nu,e}^{\rm IC} (\nu_{c,e}^{\rm IC})=(\nu/\nu_{c,e}^{\rm
IC})^{-p/2}$ and $F_{\nu,e}(\nu_{u,p}) /F_{\nu,e}(\nu_{c,e}^{\rm
IC})=(\nu_{u,p}/\nu_{c,e}^{\rm IC})^{-p/2}$, one can get
$F_{\nu,p}(\nu) / F_{\nu,e}^{\rm IC} (\nu) =
[F_{\nu,p}(\nu)/F_{\nu,p}(\nu_{u,p})]
\cdot [F_{\nu,p}(\nu_{u,p})/F_{\nu,e}(\nu_{u,p})]
\cdot [F_{\nu,e}(\nu_{u,p})/F_{\nu,e}(\nu_{c,e}^{\rm IC})] \cdot
[F_{\nu,e}(\nu_{c,e}^{\rm IC})/F_{\nu,e}^{\rm IC}(\nu_{c,e}^{\rm IC})]
\cdot [F_{\nu,e}^{\rm IC}(\nu_{c,e}^{\rm IC})/F_{\nu,e}^{\rm
IC}(\nu)]=(\epsilon_p/\epsilon_e) ^{(p-1)}$ $(\zeta_e/\zeta_p)^{(p-2)}
(m_e/m_p)^{(3p-1)/2} / [(16/3) \sigma_{T,e} \zeta_e n r
\gamma_{m,e}^{(p-1)}] \cdot (\nu_{u,p}
/\nu_{c,e}^{\rm IC})^{1/2} (\nu/\nu_{u,p})^{(p-2)/2}.$ After some
further derivations, the condition $F_{\nu,p}(\nu) > F_{\nu,e}^{\rm
IC}(\nu)$ can be translated into
\be
(1+Y_e)^{0.85}\epsilon_B > 215 k_{-1}^{-0.43} \epsilon_e^{1.02}
\epsilon_p^{-0.51}
\zeta_e^{-0.17} \zeta_p^{0.09}\alpha^{0.34}{\cal E}_{52}^{-0.22}n^{-0.35}
t_h^{-0.10} (1+z)^{0.01}\nu_{26}^{-0.04}.
\label{4}
\ee
These are the lines labeled 4 (calculated for $k=0.1$) and 4' (for
$k=1$) in Fig. 1.  In the above expression we have adopted $p=2.2$ for
both electrons and protons in order to avoid an unnecessarily
complicated expression. (The explicit spectral index $p$-dependence
can be written out straightforwardly, and is presented in the
Appendix). Notice that this criterion is mildly dependent on $\nu$,
and we have adopted $\nu=10^{26}{\rm Hz}~\nu_{26}$ for the typical
frequency in (\ref{4}).  Above this frequency (several TeV), the
self-absorption due to $\gamma-\gamma$ pair production becomes
important (\S\ref{sec:gammagamma}) and the comparison is no longer
meaningful. For lower frequencies, the constraint on the
hadron-dominant region of parameter space is more stringent.

In summary, the competition between various components in the
$\epsilon_e,
\epsilon_B$ diagram can be read off from Fig. 1. A) The first case,
applicable to frequencies $\nu<\nu_{u,e}$, is shown by the solid lines
(eqs.[\ref{1}] and [\ref{2}]), which divide the space into three
regions, in which the proton synchrotron component competes with the
electron synchrotron component, and the electron synchrotron competes
with the the electron IC component, respectively. Regions I and II are
the regions where the proton component and the electron IC component
dominate in the spectrum, respectively, while in region III, neither
of these high energy spectral components can overcome the electron
synchrotron component.  B) The second case, applicable to frequencies
$\nu >\nu_{u,e}$, is shown by the dashed lines dividing the parameter
space into two regions, in which the proton synchrotron and the
electron IC components compete with each other.  Region I' is where
the proton component as well as other hadron-related components may
overcome the electron IC emission component, while region II' is the
IC-dominated region. The dashed line 3 (eq.[\ref{3}]) is the
separation line for the case of $\nu<\nu_{c,e}^{\rm IC}$, and the
dashed line 4 (eq.[\ref{4}]) is the separation line for the case of
$\nu>\nu_{c,e}^{\rm IC}$, calculated for a typical frequency $\nu\sim
10^{26}$ Hz. For lower frequencies, this line moves leftwards, causing
the hadron-dominant phase space region to shrink.  It can be seen that
the hadron-related components are usually masked by the electron IC
component, unless $\epsilon_e$ is very small.  We note that both
eqs.(\ref{3}) and (\ref{4}) are derived under conditions which are
maximally favorable for the proton components, which includes also
adopting the extreme case of $\epsilon_p=1$, which may be an over
estimate (Vietri 1997). Bearing all these facts in mind, we expect
that the actual proton-dominated regimes (I and I') could be even
smaller than what is indicated in Fig.1. On the other hand, in both
the low energy band ($\nu<\nu_{u,e}$) and in the high energy band
($\nu>\nu_{u,e}$), the IC component is important in a much larger
portion of the parameter phase space (regions II and II').

\section{$\gamma\gamma$ pair attenuation}
\label{sec:gammagamma}

For the higher energy GRB photons in the observer frame, the
comoving frame photon energy exceeds $mc^2$, and for sufficient high
photon densities, a photon with energy $E$ in the observer frame may
be attenuated by pair production through interaction with softer
photons whose energy (also in the observer frame) is equal to or
greater than $E_{an}=(\Gamma m c^2)^2/E(1+z)^2$, depending on the
impact angle
between the two photons. This may greatly degrade the high-energy
fluence level, and the corrections due to this $\gamma-\gamma$
absorption process needs to be taken into account. The $\gamma-\gamma$
absorption in the GRB prompt phase has been studied by several
authors, e.g. Krolik \& Pier 1991; Fenimore et al. 1993; Woods \& Loeb
1995; Baring \& Harding 1997; Lithwick \& Sari 2001.  Here, however,
instead of the prompt phase we concentrate on the afterglow phase.

To treat the absorption in the afterglow phase we adopt an analytical
approach similar to the one developed for internal shocks by Lithwick
\& Sari (2001, their eq.[2]), which we adapt here to the external
shock scenario. Instead of using $\delta T$ (the temporal variation
timescale in the internal shock scenario), we use in our case an
emission timescale $t/(1+z)$, which is the expansion time as viewed by
the earth observer with the cosmological time dilation effect
correction. Assuming that the emission spectrum around $E_{an}=h
\nu_{an}$ is $L_{\nu}(\nu)=L_{\nu,0} (\nu/\nu_{an})^{-\beta}$, the
total photon number with $E>E_{an}$ can be estimated as
$N_{>E_{an}}\sim \int_{\nu_{an}}^{\infty}(L_{\nu,0}/h\nu)
(\nu/\nu_{an})^{-\beta}d\nu \cdot t/(1+z) =L_{\nu}(\nu_{an})[t/(1+z)]
/h\beta$. We adopt an averaged $\gamma\gamma$ cross section given by
$C \sigma_{T,e}$, where $C$ is a constant dependent on the photon 
spectral index. Svensson (1987) gives an analytic expression leading 
to $C=11/180$ for a photon energy density index of -1 (photon number 
density index -2), also used by Lithwick \& Sari (2001). More detailed 
calculations by, e.g., Coppi \& Blandford (1990) and B\"ottcher \& 
Schlickeiser (1997) lead to a slightly larger $C$ of the order 0.1. 
Noticing that $L_{\nu}(\nu_{an})=F_{\nu}(\nu_{an})\cdot 4\pi D^2/(1+z)$, 
the attenuation optical depth is
\be
\tau_{\gamma\gamma}(\nu)=\frac{{C}\sigma_{T,e}
N_{>E_{an}}} {4\pi [4\Gamma^2 c t/(1+z)]^2}=\frac{{C
}\sigma_{T,e} F_{\nu}(\nu_{an})D^2} {16\Gamma^4 c^2 h \beta t}~,
\label{tau}
\ee
where in terms of the quantities $t$ and $F_{\nu}(\nu_{an})$ as
measured by an Earth observer, the redshift factor has cancelled out.

It is seen from eq. (\ref{tau}) that the dominant spectral dependence
of $\tau_{\gamma\gamma}(\nu)$ is on $F_{\nu}(\nu_{an})$. For $E \sim
1$ TeV (where absorption becomes important), we find $E_{an} \sim 2.6
{\rm keV}~\Gamma_2^2(1+z)^{-2}$, which is above $h\nu_{c,e}$. In this
band, the electron synchrotron emission component dominates in a large
region of phase space (Fig.1), and even if the IC component may
potentially dominate, this happens at a later time (\S\ref{sec:results}), 
when the GeV-TeV emission is not important. Thus for the regime we are 
interested in, we can approximate $F_{\nu}(\nu_{an})= F_{\nu,max,e}
(\nu_{c,e}/\nu_{m,e})^{-(p-1)/2}(\nu_{an}/\nu_{c,e})^{-p/2}$, and take
$\beta=p/2$.  For the model of the dynamics adopted used here and for
$p=2.2$, eq.(\ref{tau}) reads then
\be
\tau_{\gamma\gamma}(\nu) \sim 0.56 C_{-1}(1+Y_e)^{-1}(1+z)^{(7p-8)/ 8}
\epsilon_B^{(p-2)/ 4}\epsilon_e^{p-1}\zeta_e^{-(p-2)}
{\cal E}_{52}^{(p+4)/ 8} n^{(p+4)/8} t_h^{(8-3p)/8}\nu_{26}^{p/2},
\ee
where $C_{-1}=C/0.1$, and the dependence of $(1+Y_e)^{-1}$ may be dropped 
for the regime I bursts ($\epsilon_e \ll \epsilon_B$ and $Y_e\sim 0$).  
Notice the mild dependence on $\nu$ and the weak dependence on $t$.  
The absorption becomes important only when $\nu$ approaches 1 TeV 
($\sim 2.4\times 10^{26}$ Hz). This simple treatment is in qualitative
agreement with the more detailed simulations of Dermer et al. 2000b
(their Fig.3).

An approximate expression for the final spectral flux including
$\gamma\gamma$ attenuation is given by the flux escaping from a
skin-depth of unit optical depth, or
\be
F_{\nu}^{ab}(\nu)=F_{\nu}^{tot}(\nu) /(1+\tau_{\gamma\gamma}),
\ee
where $F_{\nu}(\nu)^{tot}$ includes the contributions from all the
spectral components discussed above.

\section{Results and implications for high energy observations}
\label{sec:results}

With a simple numerical code that includes the three spectral
components discussed in this paper, as well as the $\gamma\gamma$
attenuation effect modeled through eq.(\ref{tau}), we have
investigated the spectra and the lightcurves in various energy bands
for different choices of the most relevant parameters, especially
$\epsilon_e$ and $\epsilon_B$.  Because of the relatively small value
of $\tau_{\gamma\gamma}$ for most energies of interest here, we have
treated this as a simple absorption process and have neglected for
simplicity the effects of the secondary pairs which it produces. The
results generally confirm the division of the $\epsilon_e ,\epsilon_B$
phase space sketched in Fig.1. Below we present some examples and
explore the detectability of the proton component and the IC component
in the afterglow phase in various bands, in particular at GeV energies
with future missions such as {\em GLAST}, and in the X-ray band with
the {\em Chandra X-ray Observatory}.  The signatures of the IC and
proton components may be detected at high energies in at least two
ways. One is through snapshot spectral fits, which require a wide and
well-sampled energy coverage, including the MeV to GeV band. Such
simultaneous measurements may be achieved in the {\em Swift} and {\em
GLAST} era. Another, simpler way, is to study the lightcurves at some
fixed high energy band, looking for a possible hardening of the
lightcurve, the details of which we discuss below.

\subsection{Detectability of the proton synchrotron component}
\label{sec:pdet}

To explore the detectability of the proton synchrotron component, we
choose as typical parameters for the regimes I and I' the values
$\epsilon_B=0.5 \epsilon_{B,.5}$, $\epsilon_e=10^{-3} \epsilon_{e,-3}$
and $n=10^2 n_2$. Notice that $\epsilon_B=0.5\epsilon_{B,.5}$,
$\epsilon_e=10^{-3} \epsilon_{e,-3}$ would fall outside region I for
the density $n\sim 1$ assumed in our Fig. 1, but adopting here a
higher density $n_2 \sim 1$ these parameters are appropriate for
region I. A denser medium will reduce $\nu_{c,e}$ (eq.[\ref{nuce}])
and consequently enlarges the region I (eq.[\ref{1}]). We do not
explore an even lower $\epsilon_e$, since it is unlikely to have
values $\epsilon_e \siml m_e/m_p \sim 0.5\times 10^{-3}$.

For the frequency range $\nu<\nu_{u,e}$, the critical time $t_p$ at
which the proton component overtakes the electron component, $t_p$,
can be derived by making eq.(\ref{Fpe}) greater than unity. From
eq.(\ref{nuce}), and taking a certain band, e.g., $\nu=10^{23}~{\rm
Hz}~\nu_{23}$, one gets
\be
t_p = 1.1 ~{\rm hr}~ \epsilon_{B,.5}^{-3}(\epsilon_{e,-3}/\epsilon_p)
^{4(p-1)} (\zeta_p/\zeta_e)^{4(p-2)} {\cal E}_{52}^{-1} n_2^{-2}
(1+z)^{-3}\nu_{23}^{-2}.
\label{tp}
\ee
The dependence on $n$ is steep, so that for a lower density medium the
overtaking time $t_p$ could be too late for observational purposes.
The dependence on the frequency is also very steep, so that the
overtaking time for slightly higher $\nu$ shifts to much shorter
time. For example, for $\nu_{23}=10$ (4 GeV), the overtaking time
moves down to $t_p \sim 40$ s.

Equation (\ref{tp}) indicates that for the above set of parameters,
the proton synchrotron component will show up in the GeV band.  The
possibility of directly detecting this component is interesting for
several reasons. One is in possibly providing a constraint on the
ratio of the proton to the electron injection fractions into the
acceleration process, $\zeta_p/\zeta_e$, which is for the interest of
the shock physics.  This would also have important implications for
the overall energetics of the fireball. A detection might also provide
information on whether the index $p$ is the same for electrons and
protons, as assumed here for simplicity (but departures from which
would be interesting).
We note, however, that the absolute luminosity of this proton GeV 
emission is too faint to be detected by {\em GLAST}, even for bursts 
at a distance of $z\sim 0.1$, unless ${\cal E}_{52}\simg 150$.
It may, however, be detectable for ${\cal E}_{52} \simg 1$ by 
future larger larger effective area ground-based GeV telescopes 
from some close regime I bursts (see \S6.4 and Fig.4).

In the X-ray band where {\em Chandra} is sensitive ($\nu\sim
\nu_{18}$), because of the steep frequency dependence $t_p \propto
\nu^{-2}$ the overtaking time occurs always too late, when the flux is
low.  Thus it appears impossible to detect the proton synchrotron
component in the X-ray band.  Figure 2a shows the snapshot spectra
including the X-ray and GeV region for the typical parameter set
representative of regime I, with a relatively dense external medium
($n_2 \sim 1$).

One remark is that the condition for the proton synchrotron component
to dominate a certain high energy band is that $\nu_{c,e}$ must
decrease with time, to make the electron component relatively less
prominent (see eq.[\ref{Fpe}]). This is not the case for the afterglow
evolution in a wind-like external medium with $n\propto r^{-2}$
(Chevalier \& Li 1999). Thus, a detection of the proton component in
the GeV band lightcurve would provide a diagnostic for an
approximately constant external medium. A non-detection, however,
would not necessarily be an argument against the constant medium,
since the phase space region for the proton component detection is
small.

\subsection{Detectability of the electron IC component}
\label{sec:icdet}

To explore the detectability of the IC component, we choose the
typical parameters for regions II and II' to be
$\epsilon_B=10^{-2}\epsilon_{B,-2}$, $\epsilon_e=0.5 \epsilon_{e,.5}$
and $n=1$, similarly to Sari \& Esin (2001).  Following the same
procedure to derive (\ref{tp}), one can derive the critical time
$t_{IC}$ when the IC component overtakes the synchrotron component at
a typical frequency $\nu<\nu_{u,e}$.

In the X-ray band, the overtaking time of the IC component usually
occurs in the slow-cooling regime, which we will assume in the
following discussions (see below).  The crossing point between the
synchrotron spectral component and the IC spectral component,
$\nu_e^{\rm IC}$ (Sari \& Esin 2001), could in principle be either
above or below $\nu_{m,e}^{\rm IC}$.  If $\nu_e^{\rm
IC}>\nu_{m,e}^{\rm IC}$, eq.(\ref{FIC1}) could be used directly, and
the overtaking condition is $(16/ 3) \sigma_{T,e}\zeta_e nr
\gamma_{m,e}^{(p-1)} \left({\nu /
\nu_{c,e}}\right)^{1/2}>1$. The complication in comparing the regions I
and I' is that the Compton cooling factor $(1+Y_e)$ in the expression
of $\nu_{c,e}$ (eq.[\ref{nuce}]) can no longer be neglected.  For
$(\eta \epsilon_e/\epsilon_B)^{1/2}\gg 1$, one has $(1+Y_e) \simeq
(\eta \epsilon_e/\epsilon_B)^{1/2}$, where in the slow cooling phase
$\eta=(\gamma_{m,e} /\gamma_{c,e})^{p-2}$. We have derived the
overtaking time for this case, where $t_{\rm IC} \propto
\nu^{[4(4-p)/(3 p^2-23p+36)]}$ (cf. Sari \& Esin
2001). For reasonable values of $p$ (e.g. 2.2-2.4), the quantity
$(3p^2-23p+36)$ is close to zero, causing a very sharp, and probably
unphysical, dependence of $t_{\rm IC}$ on all parameters.  However,
the value of $t_{\rm IC}$ derived in this manner is not generally
useful for our purposes here. The reason is that, to ensure
$\nu_e^{\rm IC}>
\nu_{m,e}^{\rm IC}$, the $\epsilon_e$ and $\epsilon_B$ should be close
to the values near the boundary between the regions II and III
(eq.[\ref{2}]), so that both values are comparable. In such a regime,
the approximation of $(1+Y_e) \simeq (\eta
\epsilon_e/\epsilon_B)^{1/2}$ no longer holds, and one cannot get a
simple analytic expression for $t_{\rm IC}$.

More generally, the overtaking occurs when $\nu_e^{\rm
IC}<\nu_{m,e}^{\rm IC}$ in situations where $\epsilon_e\gg
\epsilon_B$, as is the case for the typical values adopted above (and
in Sari \& Esin 2001). Noticing that in this regime $F_{\nu,e}^{\rm
IC} (\nu)=F_{\nu,max,e}^{\rm IC}(\nu/\nu_{m,e}^{\rm IC})^{1/3}
=F_{\nu,max,e}^{\rm IC}(\nu/\nu_{m,e}^{\rm IC})^{-(p-1)/2}
(\nu/\nu_{m,e}^{\rm IC})^{(3p-1)/6}$, using eq.(\ref{FIC1}) the
overtaking condition can be written as $(16/ 3) \sigma_{T,e}\zeta_e nr
\gamma_{m,e}^{(p-1)}
\left({\nu / \nu_{c,e}}\right)^{1/2}(\nu/\nu_{m,e}^{\rm
IC})^{(3p-1)/6}>1$. This finally leads to
\be
t_{\rm IC}=3.4~{\rm days}~\epsilon_{e,.5}^{0.89}\epsilon_{B,-2}^{0.08}
\zeta_e^{1.63} {\cal E}_{52}^{-0.06} n^{-0.66} (1+z)^{-0.36}
\nu_{18}^{-0.68}
\label{tIC}
\ee
for $p=2.2$ and $\nu=10^{18} {\rm Hz}~\nu_{18}$ (X-ray band).  The
$p$-dependence in the above expression is more cumbersome, and is
given in the Appendix. This result is in general agreement with Sari
\& Esin (2001)\footnote{The numerical indices in the equation
(\ref{tIC}) are in agreement with Sari \& Esin's results for $p=2.2$,
except the index for $(1+z)$, where we have included an additional
factor from the frequency redshift correction, i.e., -0.36=0.32-0.68.
}, and in addition here we have explicitly presented the $\zeta_e$,
${\cal E}_{52}$, and $\nu$ dependences which are absent in their
paper. A roughly factor of two difference on the overtaking time
($\sim 7.7$ days in their case) may be caused by slightly different
coefficients adopted in both works for the $\nu_{m,e}$, $\nu_{c,e}$,
$F_{\nu,max,e}$, etc. Nonetheless, this confirms Sari \& Esin's
finding that in a reasonably dense medium, the IC component can be
directly detected by {\em Chandra} a couple of days after the burst
trigger.  We note that a substantial flattening of the X-ray light
curve for GRB 000926 has been detected by {\em Chandra} (Piro et
al. 2001). Since the proton component cannot show up in the X-ray band
under any circumstances, as we argued in \S6.1, such a flattening may
indicate a direct detection of the IC emission of the
electrons \footnote{After this paper was submitted, we noticed that
Harrison et al (2001) have performed a detailed fit to the snapshot
spectra of the afterglow data of this burst at 2 day and 10 day. The
X-ray emission data are consistent with an IC emission component. The
best-fit parameters they derived lie in our Regime II. This agrees
with the discussions presented here.}. An
alternative interpretation is advanced in Piro et al. (2001).

Extensive efforts have been made to determine key fireball parameters
such as ${\cal E}, \epsilon_B, \epsilon_e, n$ using snapshot spectral
fits extending from radio to X-rays on well studied GRB afterglows
(e.g. Galama et al. 1998; Wijers \& Galama 1999; Panaitescu \& Kumar
2001). A cautionary point which needs to be stressed
about these analyses is that the spectrum which is observed need not
be, as is generally assumed, solely due to electron synchrotron
radiation, especially when using late time ($\simg$ a couple of days)
data in the fitting.  According to our results in this paper, as long
as the $\epsilon_e,\epsilon_B$ phase space is in the regions I or III,
the standard fitting assumption (i.e. electron synchrotron dominance)
is safe, since there are no high energy (proton or IC) spectral
components appearing in the X-ray band.  However, in the region II of
parameter space (which includes values of $\epsilon_B,\epsilon_e$
often derived from such fits), the analysis may not be
self-consistent, since the X-ray data points may be due the IC
component. This caution applies to even earlier snapshot spectral
fits, if the burst happens to occur in a denser medium (notice the
negative dependence of $n$ on the $-(2/3)(5p-26)/(3p^2-8p-12)$ index).

The negative dependence on $\nu_{18}$ of $t_{\rm IC}$ (the explicit
index is $-(2/3)(3p+2)(p-4)/(3p^2-8p-12)$, see also Sari \& Esin 2001)
indicates that for energy bands above X-rays, the overtaking time is
much earlier\footnote{At higher energy bands, $\nu_{\rm
IC}<\nu_{m,e}^{\rm IC}$ may be no longer satisfied, and the overtaking
time may occur in the fast-cooling regime. Nonetheless, the negative
dependence on $\nu$ of $t_{\rm IC}$ generally holds.}. In fact, in the
GeV band, the IC component dominates almost throughout the entire
afterglow phase.  For the typical parameter set corresponding to the
IC dominated region II, Figure 2b shows the time evolution of the
snapshot spectra. For completeness, we present also in Figure 2a the
snapshot spectra for the typical parameter set in the proton dominated
region I, while Figure 2c shows the snapshot spectra for parameters in
the electron synchrotron dominated region III. We can see that in this
latter region, neither of the two high energy components is prominent
below $\sim$ GeV energies (although at $\simg$ GeV there is a lower
level IC component).  Two explanations ought to be made about our
code. First, to avoid adding up an unphysical component in
the electron synchrotron self-absorption band, we have arbitrarily
defined the self-absorption cut-off in the proton component. The
self-absorption segment in the electron IC component is still plotted
with a slope 2 rather than 1 (Sari \& Esin 2001) for the convenience
of code developing, which does not influence the final broad-band
spectrum. Second, at the cut-off frequency of each component, we have
adopted a sharp cut-off while a more realistic cut-off should be
exponential. The same applies for the sharp jumps in the lightcurves
presented in Figs.3 and 4.

In Figure 3, we present the X-ray lightcurves for the typical bursts
in the three different regimes. While the regime I and III afterglows
show a monotonous decay in this band, we show that the regime II
afterglows can show interesting bump features, due to the dominance of
the IC component at a later time, in qualitative agreement with
Panaitescu \& Kumar (2000) and Sari \& Esin (2001).

\subsection{GeV afterglows and GRB 940217}
\label{sec:gevdet}

The extended GeV emission 1.5 hours after the trigger of GRB 940217
detected by {\em EGRET} (Hurley et al. 1994) indicates that a high
energy spectral component can extend into the GeV band for a long
period of time, at least in some bursts. In principle, this could be
either due to the proton synchrotron emission in the regime I, or due
to the electron IC emission in the regime II, or even due to the
electron synchrotron emission in the regime III. We will show below
that the regime II IC-dominated origin is the more plausible
explanation.

The peak energy flux expected from various scenarios can be estimated
straightforwardly. For the proton synchrotron in regime I, we have
\begin{eqnarray}
\nu F_{\nu,p} ({\rm GeV})  =
\nu F_{\nu,max,p} (\nu /\nu_{m,p})^{-(p-1)/2} \sim
 1.4\times 10^{-14} {\rm ergs~s^{-1} ~cm^{-2}} & \nonumber \\ \times
 \epsilon_p^{(p-1)} \zeta_p^{(2-p)} \epsilon_{B,.5}^{(p+1)/4}
 n_2^{1/2} {\cal E}_{52}^{(p+3)/4} D_{28}^{-2} (1+z)^{(9-p)/4} &
 t_h^{-3(p-1)/4} \nu_{23}^{(3-p)/2}
\label{nuFnup}
\end{eqnarray}
(see also Fig.2). This is more than one order of magnitude below the
calculated level of B\"ottcher \& Dermer (1998) as well as the
analytical estimate of Totani (1998). The main discrepancy with both
of these results is due to their having adopted $\gamma_{m,p}=\Gamma$,
rather than the more accurate lower value we adopted in equation
(\ref{gammam}), and also due to the fact that we calculate the flux
coefficient using $p=2.2$ while they use $p=2$, which further enhances
the discrepancy between $\gamma_{m,p}$ and $\Gamma$, and which gives a
milder $\nu$-dependence. To test this, we have substituted
$\gamma_{m,p}=\Gamma$ and $p=2$ in our code, and this reproduces
B\"ottcher \& Dermer's results. We conclude that the correction
introduced by using here the more realistic $\gamma_{m,p}$
(eq.[\ref{gammam}]) is essential, and that the previous rough
estimates using $\gamma_{m,p}=\Gamma$ can considerably overestimate
the proton synchrotron flux level. Another feature, which can be seen
from equation (\ref{nuFnup}), is the negative temporal decay of the
flux level with the index $-3(p-1)/4 \sim -0.9$, which indicates that
even if the the proton synchrotron emission flux level is detectable
in the early afterglow phase, it will drop with time as time goes by.
For the regime III, electron-synchrotron-dominated case, the trend is
similar, with a steeper temporal index $-(3p-2)/4 \sim -1.15$ (Fig.4).

In the IC-dominated regime II, contrary to the proton component, the
IC component itself has a bump peaking at $\nu_{m,e}^{\rm IC}$ (for
slow-cooling) or $\nu_{c,e}^{\rm IC}$ (for fast-cooling) in the
$F_\nu$ plot.  The peak will sweep the band $\nu=10^{23}{\rm
Hz}~\nu_{23}$ at $t=t_{max} \sim 0.3 {\rm hr}~ \epsilon_{e,.5}^{16/9}
\zeta_e^{-16/9}
\epsilon_{B,-2}^{2/9} {\cal E}_{52}^{1/3} n^{-1/9} (1+z)^{5/9}
\nu_{23}^{-4/9}$ in the slow-cooling regime.
The temporal index before the flux reaches its peak is 1
(slow-cooling) or $(8-3p)/3(4-p) \sim 0.26$ (fast cooling), and is
$(11-9p)/8 \sim -1.1$ (slow cooling, which is usually the case) after
the flux has passed its peak.  The peak energy flux can be estimated
as $\nu F_{\nu,max,e}^{\rm IC} (t=t_{max})
\sim 3.0\times 10^{-10} {\rm ergs~ s^{-1}~cm^{-2}}$ for $\epsilon_B \sim
0.01$. This temporal evolution is mild, which allows
a substantial GeV emission component lasting hours after the GRB
trigger.

The {\em EGRET} flux sensitivity above 100 MeV is $\sim 10^{-7} {\rm ph
~s^{-1}~cm^{-2}}$ for point-source observations over a period of two
weeks in directions away from the Galactic Plane. Correcting for an
average effective on-source observing fraction of 45\% (D. J. Thompson, 
private communication), the fluence threshold may be estimated as 
$\sim 10^{-7} \cdot T (t/T)^{1/2} \sim 5\times 10^{-8}t^{1/2}~
{\rm ergs~cm^{-2}}$ at an average energy 400 MeV, where 
$T=14\cdot 86400 \cdot 45\%$, and $t$ is integration time in seconds.
This fluence sensitivity may be extrapolated down to integration times 
such that at least, say, 5 photons are collected. For even shorter 
integration times, the flux sensitivity may be defined, e.g. by the 
criterion that at least 5 photons are collected, which is 
$\sim 5/(A_{eff} t)~ {\rm ph~s^{-1}~cm^{-2}}$. For {\em EGRET},
$A_{eff} \sim 1500 {\rm cm}^2$, giving an estimated fluence threshold 
in the low integration time regime of $\sim 5/1500 \cdot 400 {\rm MeV}
\sim 2.1\times 10^{-6} {\rm ergs~cm^{-2}}$. As shown in Figure 4,
for ${\cal E}_{52}\sim 1$, the {\em EGRET} fluence threshold level is 
not reached by regime I or III bursts, but this level is attainable for 
a regime II burst located at closer distances (e.g. $z=0.1$), or in a 
higher density environment. Thus the late ($\sim$ hour) GRB 940217 
afterglow was most likely dominated by the electron IC emission from a 
nearby or dense-medium regime II burst. This agrees with \Mesz~ \& Rees 
(1994), and Dermer et al. (2000b) drew a similar conclusion by detailed
simulations using a specific set of parameters, i.e., $\epsilon_e=0.5$, 
$\epsilon_B \leq 10^{-4}$, which lies in our regime II. 
The Gamma-ray Large Area Space Telescope ({\em GLAST})
currently under construction will have a flux sensitivity of $\sim 1.6
\times 10^{-12} {\rm ergs~s^{-1}~cm^{-2}}$ for long term observations
(Gehrels \& Michelson 1999), roughly 40 times more sensitive than 
{\em EGRET} in the point-mode. In the low flux regime, given the
effective area of $\sim 8000 {\rm cm}^2$ (Gehrels \& Michelson 1999),
it is only about a factor of 5 more sensitive than {\em EGRET}. The
fluence threshold for {\em GLAST} is roughly $\sim 1.2\times
10^{-9}t^{1/2} ~{\rm ergs~cm^{-2}}$ for long integration time regime,
and $\sim 4.0\times 10^{-8} {\rm ergs~cm^{-2}}$ for short integration
time regime (again assuming that at least 5 photons are collected). This
will make most regime II burst afterglows detectable at a typical
cosmological distance and in a moderate density medium (Fig.4).

In Figure 4 we show the GeV lightcurves for bursts typical of the
three different parameter regimes and ${\cal E}_{52}=1$. For comparison 
with future observations, we have integrated over the 400 MeV - 200 GeV 
band to get the total energy fluence {\em GLAST} can collect during a 
certain time duration $t$. The sensitivity threshold of {\em EGRET} and 
{\em GLAST} are indicated. One sees that regime II burst afterglows would
be generally detectable by {\em GLAST} within hours after the burst
trigger. Bursts in regime I and III are generally non-detectable by
{\em GLAST}, even for burst at $z=0.1$. Increasing total energy budget
(${\cal E}_{52}$) or ambient density ($n$) can increase the
detectability of these bursts. 
Since for transient events, the key
factor of sensitivity is the effective collecting area, some future
ground-based larger area GeV telescopes, such as the $5@5$ - 5 GeV
energy threshold array of imaging atmospheric Cherenkov telescopes at
5 km altitude (Aharonian et al. 2001), may have better chance to
detect the regime I and III bursts (at the prompt phase), and of
course, to collect more photons from regime II bursts.
A criterion to differentiate between nearby regime I and
III bursts is that the lightcurve for the regime I burst is flatter.
In any case, we conclude that an extended GeV afterglow is a
diagnostic of a regime II (IC-dominated) burst.

Recently, an energy flux upper limit $J(E > 760 {\rm GeV}) < 9.4\times
10^{-12} {\rm ergs~ cm^{-2} ~s^{-1}}$ in GRB 010222 was obtained in a
4-hour measurement with the stereoscopic {\em HEGRA} Cherenkov
telescope system, 19 hours after the burst trigger (Goetting
\& Horns 2001). Given the cosmological distance of $z\geq 1.477$ (Jha
et al. 2001), this is consistent with our model prediction in this
paper (see Fig.2), even for the most favorable regime II case
(Fig.2b).

\subsection{Prospects from broadband observations in the
{\em Swift}-{\em GLAST} era}
\label{sec:swift-glast}

The {\em GLAST} mission will be launched in 2005, with a sensitivity
range in the 20 MeV-300 GeV, complemented by the Glast Burst Monitor
({\em BGM}) whose energy range extends from a few keV to 30
GeV. Another broad band GRB mission, {\em Swift}, will be launched in
2003, and will be sensitive in the optical, X-ray and $\gamma$-rays up
to $\siml$ 140 keV. At the same time, ground-based experiments such as
{\em Milagro}, {\em HESS}, {\em Veritas}, {\em MAGIC} and {\em
Cangaroo-III} may provide $\simg 0.5$ TeV, or in some cases $\simg 30$
GeV data or upper limits.  In the {\em Swift}-{\em GLAST} era,
simultaneous broad-band observations at the very earliest stages of
the GRB afterglows will become possible, which will bring invaluable
information about GRB shock physics and the central engine.  Here we
note several interesting issues which can be addressed in the {\em
Swift}-{\em GLAST} era:

1. Sari \& Esin (2001) pointed out that due to the IC cooling, there
are two possible solutions of the unknown fireball and shock
parameters for a same set of observables from the low-energy afterglow
fits. For high density mediums, these two scenarios may be
distinguished from the late-time X-ray observations. But for low
density mediums, the two solutions are degenerate and
indistinguishable with the present data. However, we note that the two
sets of solutions lie within the regime III and regime II,
respectively. This provides a natural way to distinguish between the
two scenarios by using the GeV afterglow data. If an extended GeV
afterglow is detected by {\em GLAST}, then the parameter space should
be in regime II where the IC
component dominates in the high energy band. In this case eqs.(4.17)
to (4.20) of Sari \& Esin (2001) will apply. Otherwise, the parameter
space should be in regime III, where no prominent emission component
shows up in the GeV band. This is the case that Sari \& Esin's
eqs.(4.13) to (4.16) may apply.

2. It has been proposed by Waxman (1995) and Vietri (1995) that GRBs
are likely sites to produce ultra-high-energy cosmic rays (UHECRs),
and that the $10^{20}$ eV excess UHECRs detected are of the GRB
origin.  This hypothesis is subject to debate (cf. Stecker 2000;
Mannheim 2000; Scully \& Stecker 2001). We note that future GeV
observations with {\em GLAST} may be able to add important criteria on
the debate, at
least for the external shock scenario.  To accelerate protons to
ultra-high energies around $10^{20}$ eV, $\epsilon_B$ must be close to
unity (Waxman 1995; Vietri 1995; Rachen \& \Mesz~ 1998, but see Dermer
2001). If substantial
extended GeV afterglows are common among GRBs, and if the redshift
measurements from the low frequency afterglow observations indicate
that the GRBs are around $z\sim 1$, this will impose severe
constraints on the UHECR acceleration theory by the external shock
scenario, since the regime II generally favors a small $\epsilon_B$.
(However, our calculations do not apply to a possible UHECR
acceleration in internal shocks).  On the other hand, if long duration
GeV afterglows are not common, and for a few nearby bursts a GeV
(prompt) lightcurve hardening is detected by {\em GLAST} or some other
future telescopes, this could be attributable
to proton synchrotron emission, in which case both $\epsilon_p$ and
$\epsilon_B$ are close to unity. This would provide support to the
theory of UHECR origin in GRBs. Since these components are not masked
by the electron synchrotron and the IC components, this would also
hint at a small $\epsilon_e$ (e.g. due to a weak coupling between
electrons and protons, as argued by Totani 1998; 2000).

3. Although most present snapshot spectral fits assume that the
unknown shock parameters ($\epsilon_e$, $\epsilon_p$, $\epsilon_B$)
are constant with time, there is no a priori reason for this simplest
assumption. Evolutions of one or more of these parameters are in
principle possible (e.g. Dermer et al. 2000b). Present observations
are too crude to explore these possibilities, but future broadband
observations will provide the opportunity to explore this important
issue. In our Figure 1, we present generically the parameter space of
various regimes as well as their dependences on some other parameters
including the observation time. In the whole discussions in \S6, we
have assumed the constant equipartition parameters. If there exist
substantial evolutions of these parameters, the blastwave parameters
will change in the $\epsilon_e-\epsilon_B$ space, and may, in some
cases, switch the regime they belong to as time goes by. This will
bring some additional interesting signatures in the lightcurves in
various bands, and further more detailed studies may provide
diagnostics on the possible evolutional effects.

\section{Summary}
\label{sec:sum}

We have studied GRB afterglow snapshot spectra and lightcurves over a
broader band than usual by including the canonical electron
synchrotron emission as well as two other high energy spectral
components, i.e., the proton synchrotron emission component and the
electron synchrotron self-inverse-Compton emission component. We have
in particular concentrated upon the X-ray to GeV-TeV ranges, including
the effect of attenuation by $\gamma\gamma$ pair formation.  This
investigation has the advantage, relative to prior ones, of bringing
together in a single coherent treatment the effects of these various
high energy mechanisms, which hitherto had been mostly treated singly
or in twos, within the context of a specific GRB afterglow dynamical
model.  This is carried out over a wider range of parameter phase
space than hitherto, to allow a global view on the relative importance
of the various spectral components.

For the frequency range below the electron's synchrotron cut-off,
$\nu<\nu_{u,e}$, there is a competition between the electron
synchrotron component on the one hand, and the proton synchrotron
component or the electron IC component on the other, which can affect
the higher energy bands including X-rays or above. This competition
divides the $\epsilon_e,\epsilon_B$ phase space into three regimes
(Fig.1). We have explored the range of validity of these regimes, and
discussed the conditions for which these high energy spectral
components would show up in various bands, especially in the GeV and
the X-ray band. The conclusion is that the IC component is likely to
be important in a relatively large region of parameter space, while
the conditions for which the proton synchrotron component is important
involve a small, but non-negligible, region of parameter space.  One
interesting consequence is that there is a substantial region (regime
III) in which neither of the two high energy components are important.
Above the electron synchrotron cut-off, the competition is
between the electron IC component and the hadron-related photo-meson
decay components, which we treated as a reduced extension of the
proton synchrotron component. Again, the phase space region where the
latter effects are important in the afterglow is small.  We also find
that for the external shock and the afterglow phase the
$\gamma-\gamma$ absorption is not important below the TeV range.

A general conclusion is that the most likely origin for an extended
high energy afterglow component at GeV energies is from the electron
IC component. Not only is the phase space region where the IC component
dominates (II and II') much larger than that where the proton
component dominates (I and I', see Fig. 1), but its intensity is also
much higher than that of the hadron components, and the time scale
during which an appreciable flux level is maintained in the GeV band
is much longer than for the hadronic components.  In the parameter
regime favorable for the IC emission, this component is observable at
and above the X-ray band. In the X-ray band, it will lead to a
flattening of the light curve at late times, as long as the medium
density is not too low (see also Sari \& Esin 2001). Above the X-ray
band, the time after which IC emission becomes dominant appears
earlier, and the IC component dominates the GeV-band emission almost
from the onset of the afterglow phase.  In general, a high external
density medium favors the detectability of the IC component. Such an
IC component is likely to have been responsible for the GeV photons
detected from GRB 940217 with {\em EGRET}, and similar events should
in the future be detectable by {\em GLAST}.

The proton synchrotron component, as well as the hadron-related
photo-meson electromagnetic components in the afterglow radiation are
likely to be, in most cases, less important than previous approximate
estimates indicated (Vietri 1997; B\"ottcher \& Dermer 1998; Totani
1998; Totani 2000).  The $\epsilon_e,\epsilon_B$ phase space region
where this component overcomes the electron synchrotron and the IC
components is very small, unless the medium density is high. Even in
the most favorable region of the phase space, the proton component is
not expected to show up in the X-ray band, but it may overcome the
electron synchrotron component in the GeV band shortly after the onset
of the afterglow phase, if the medium density is moderately high. 
However, the flux level is low and drops with time, and is generally 
not detectable by {\em GLAST} unless for some extreme bright bursts 
at close enough distances. We note also that the value ${\cal E}_{52}=1$ 
used for the estimates in the figures is per steradian, and narrow jets 
could in principle give values $\simg 10^2$. The detection of a proton 
component would be of extremely high interest (\S \ref{sec:pdet}) for 
the GRB and shock physics.  Its detection, if successful, would imply,
depending on its strength, fireball and shock parameters which are
more extreme than currently commonly assumed. In particular, it would
provide a diagnostic for a high $\epsilon_p,\epsilon_B$ and/or a low
$\epsilon_e$.

We point out a simple way to break the current parameter space
degeneracy which currently, from low frequency observations alone
(X-rays and below), prevent the unambiguous determination of the
unknown fireball and shock parameters, through the use of snapshot
spectral measurements extending into the GeV range. We also suggest a
way to test the hypothesis of a GRB origin for UHECR using combined
{\em Swift} and {\em GLAST} data. Such observations may also provide
diagnostics for the presence of a quasi-homogeneous versus a wind-like
inhomogeneous external medium.

Finally, we note that the condition for the spectral IC component to
be prominent (eq.[\ref{2}] and Fig.1) usually covers the parameter
regime in which IC cooling is important ($\eta\epsilon_e /\epsilon_B
>1$, Sari \& Esin 2001). Thus, the current snapshot spectral fits
ought to be made with caution for the X-ray data points when the IC
cooling is important, especially for the data at later times, and for
cases which may involve a high external medium density.

\acknowledgments{We are grateful to C. D. Dermer, D. J. Thompson and
M. B\"ottcher for informative correspondence, to the referee for valuable
comments, and to NASA NAG5-9192 and NAG5-9193 for support.}

\appendix{
\section{Explicit $p$-dependences of equations (18) and (23)}

In this appendix, we explicitly present the $p$-dependent indices in
equations (18) and (23). For equation (18):
\begin{eqnarray}
(1+Y_e)^{8\over 16-3p}\epsilon_B & > 215 k_{-1}^{-{4\over 16-3p}}
\epsilon_e^{8(p-1)\over 16-3p}\epsilon_p^{-{4(p-1)\over 16-3p}}
\zeta_e^{-{8(p-2)\over 16-3p}} \zeta_p^{4(p-2)\over 16-3p}
\alpha^{4(3-p)\over 16-3p} \nonumber \\
& \times{\cal E}_{52}^{-{13-4p\over 2(16-3p)}} n^{-{11-2p\over
2(16-3p)}} t_h^{-{4p-7\over 2(16-3p)}} (1+z)^{9-4p\over
2(16-3p)}\nu_{26}^{-{2(p-2)\over 16-3p}}~.
\end{eqnarray}
For equation (23):
\begin{eqnarray}
t_{\rm IC}&=3.4~{\rm
days}~\epsilon_{e,.5}^{4(3p^2-8p-7)\over3(3p^2-8p-12)}
\epsilon_{B,-2}^{3p^2-10p+4\over 3(3p^2-8p-12)}
\zeta_e^{4(3p^2-5p-22)\over 3(3p^2-8p-12)} \nonumber \\
&\times{\cal E}_{52}^{{2p^2-4p+1\over 2(3p^2-8p-12)}}
n^{-{2(5p-26)\over 3(3p^2-8p-12)}} (1+z)^{-{3p^2-16p+4\over
3(3p^2-8p-12)}}
\nu_{18}^{-{2(3p+2)(p-4)\over 3(3p^2-8p-12)}} ~.
\end{eqnarray}
}

\newpage

\begin{figure*}
\centerline{\psfig{file=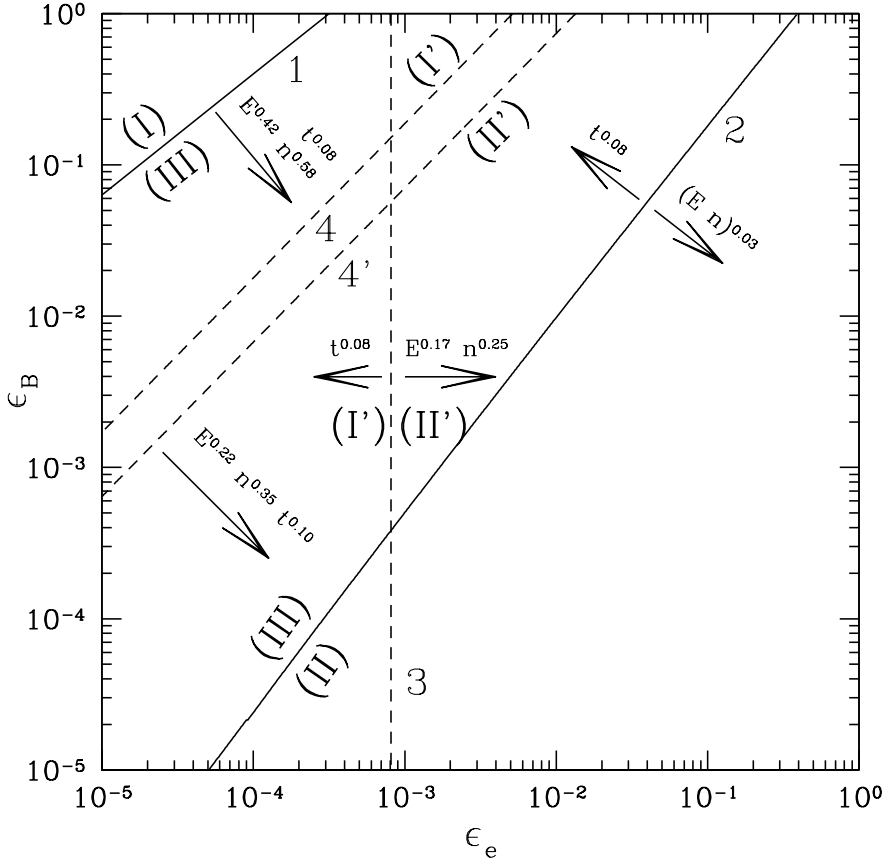,width=12.0cm,height=12.0cm}}
\caption{Regions in the $\epsilon_e$,$\epsilon_B$ parameter space
where the various radiation mechanisms dominate at selected
frequencies. A) For photon energies $\nu <\nu_{u,e}$ (the synchrotron
frequency for electrons at the upper end of their energy distribution)
the solid lines 1 and 2 divide the space into three regimes. Regime I
is where the proton synchrotron component overcomes the electron
synchrotron component; regime II is where the electron IC component
overcomes the electron synchrotron component; and regime III is where
the electron synchrotron component dominates the other two. B) For a
higher energy band with $\nu>\nu_{u,e}$, the space is divided into two
regimes I and I' by a dashed line 3, or a dashed line 4 (4'), depending
on the subcase. Regime I' is where the proton-related components
overcome the electron IC component, and regime II' is where electron
IC dominates over the proton components. For $\nu<\nu_{c,e}^{\rm IC}$
(the inverse Compton-boosted frequency of synchrotron photons radiated
by electrons at the cooling break energy) the separation is given by
line 3, which does not depend on the frequency. For
$\nu>\nu_{c,e}^{\rm IC}$ the separation line is frequency-dependent,
and given by line 4 (4'), which are drawn for $\nu\sim 10^{26}$
Hz. Line 4 assumes the reduction factor is $k=0.1$, while line 4'
assumes $k=1$ (see text). The dependences of the separation lines on
$t$, $n$ and ${\cal E}$ are indicated on the plot, which cause the
different regimes to enlarge or shrink with these parameters. All
lines are drawn using the following parameters: $\xi_e=\xi_p=1$,
$\epsilon_p=1$, ${\cal E}_{52}=1$, $n=1$, $\alpha=1$, $z=1$, $p=2.2$
and $t=1$ hr. For a flatter $p$ (close to 2), the lines 1, 2, 4 and 4'
are flatter, and the regions I and I' will be slightly enlarged. For a
steeper $p$, the trend is the opposite.}
\end{figure*}

\begin{figure*}
\centerline{\psfig{file=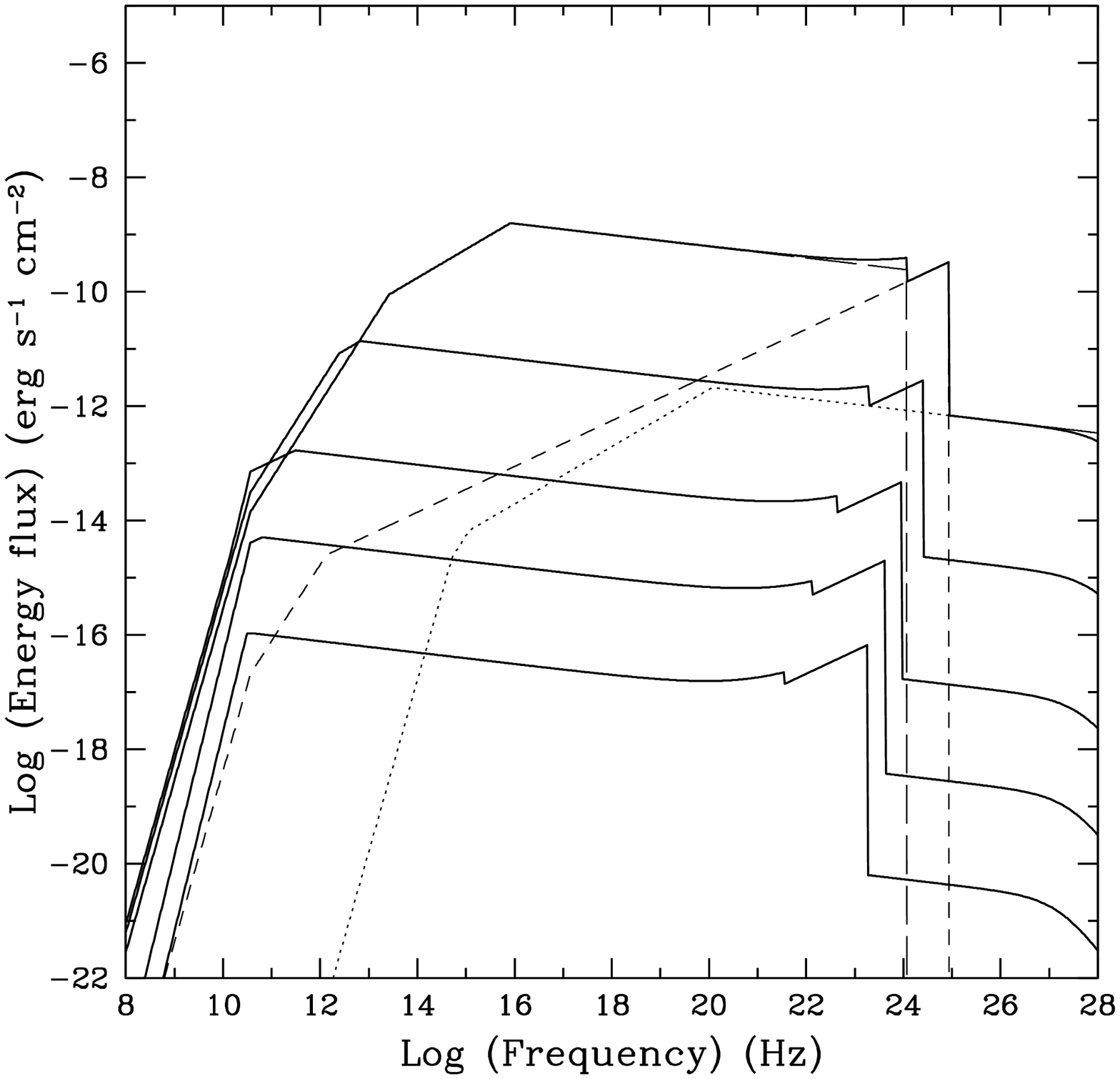,width=7.0cm,height=7.0cm}
\psfig{file=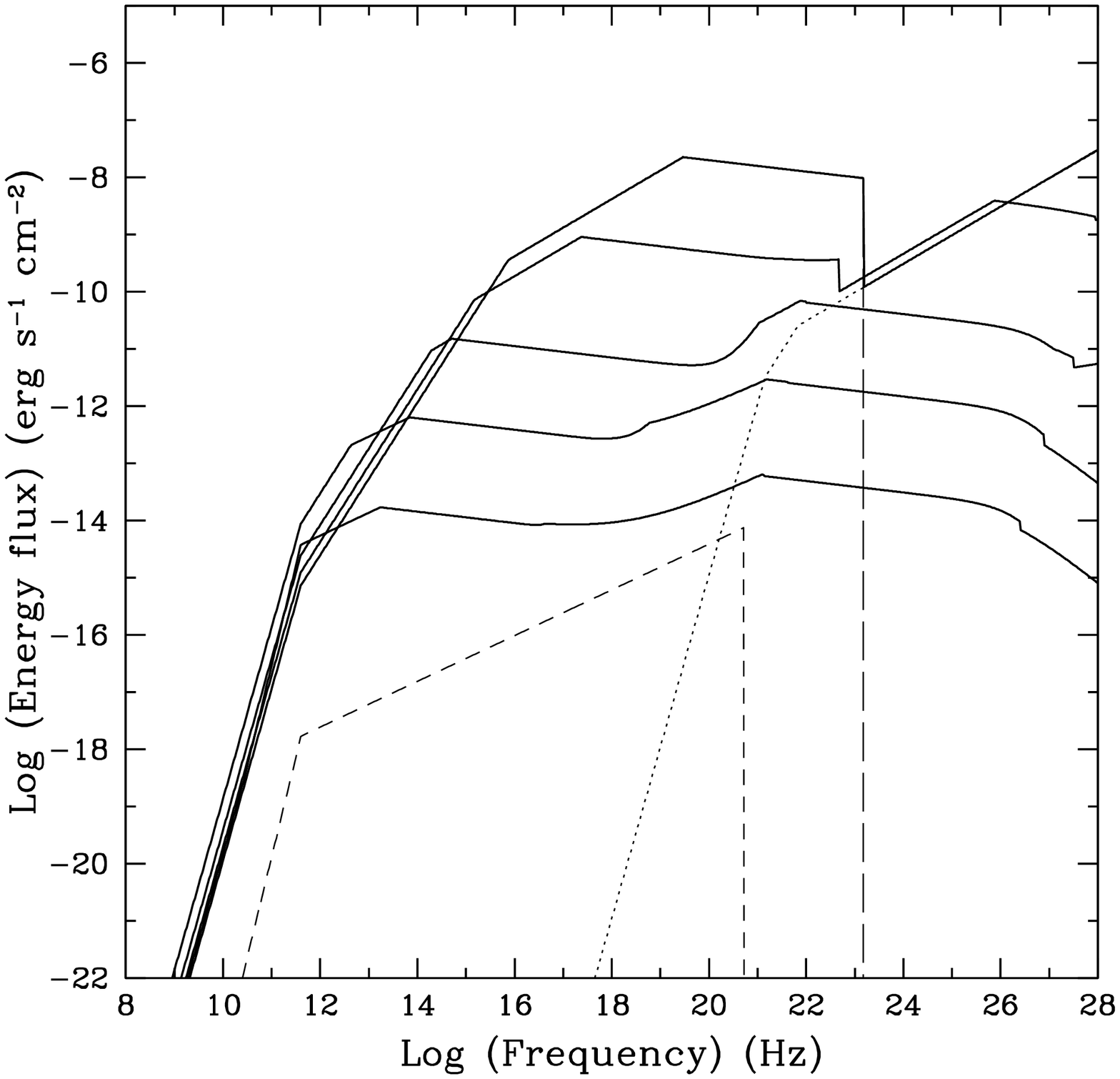,width=7.0cm,height=7.0cm}}
\centerline{\psfig{file=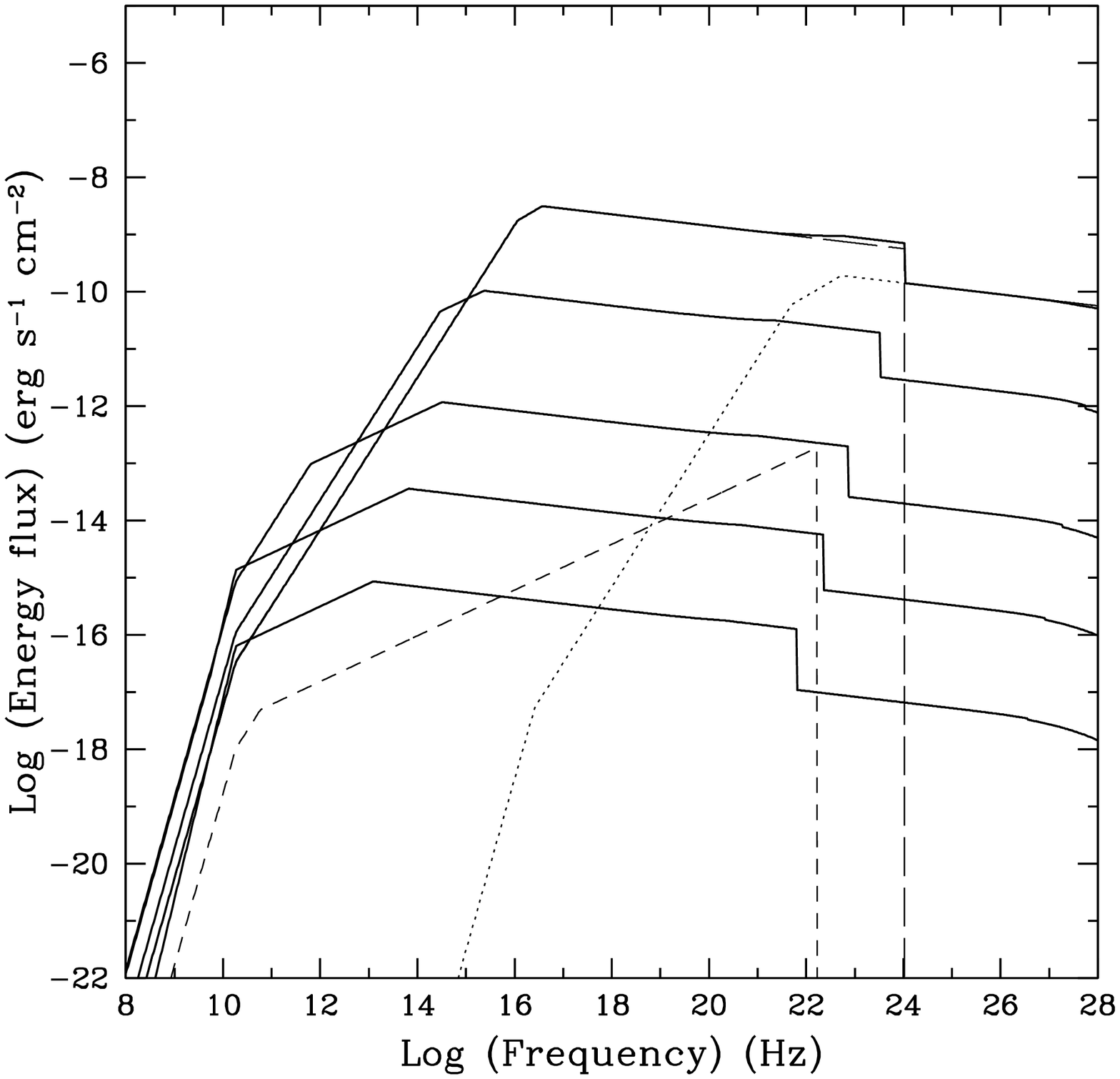,width=7.0cm,height=7.0cm}}
\caption{Temporal evolution of the broad-band spectra of GRBs.
Thick solid curves are the final spectra for various observer times,
starting from (top) the onset of the afterglow, 1 minute, 1 hour, 1
day to (bottom) 1 month, respectively.  The sharpness of the breaks
and cutoffs is an artifact of the analytical approximations, in
reality these would be smoother transitions.  For the top curve,
contributions from the various radiation components are also
plotted. Long dashed are electron synchrotron, short dashed are proton
synchrotron, and dotted lines are electron inverse Compton
emission. The thin solid line is the total energy flux level without
$\gamma\gamma$ absorption correction, while the thick solid line is
the energy flux level after the $\gamma-\gamma$ self-absorption
correction. The intergalactic absorption, which also becomes important
around $\nu=10^{26}$, is distance dependent and has not been included
in this graph. Here all plots are calculated for standard parameters
$z=1$ (flat, $\Lambda=0$ universe), $\zeta_e=\zeta_p=1$,
$\epsilon_p=1$, ${\cal E}_{52}=1$, $\alpha=1$, $z=1$, $p=2.2$, and
$\Gamma_0=300$, while $\epsilon_e$, $\epsilon_B$ and $n$  vary for
the different regimes.  (a) A typical regime I burst:
$\epsilon_e=10^{-3}$, $\epsilon_B=0.5$, $n=100 ~{\rm cm^{-3}}$. (b) A
typical regime II burst: $\epsilon_e=0.5$, $\epsilon_B=0.01$, $n=1~
{\rm cm^{-3}}$.  (c) A typical regime III burst: $\epsilon_e=0.01$,
$\epsilon_B=0.1$, $n=1~ {\rm cm^{-3}}$.}
\end{figure*}

\begin{figure*}
\centerline{\psfig{file=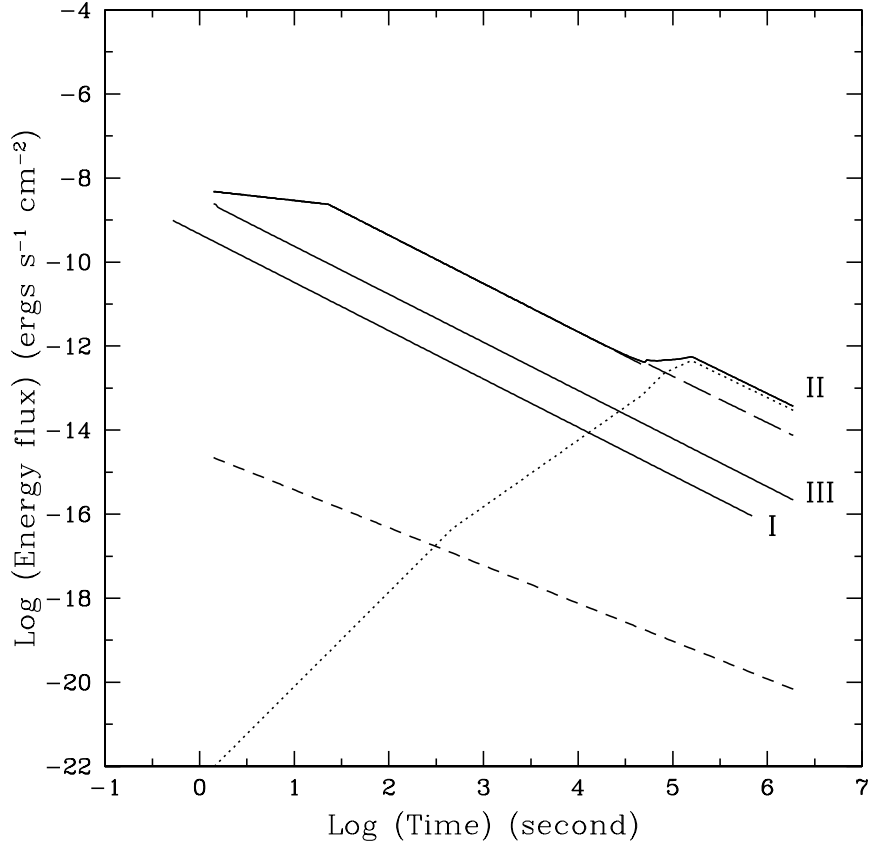,width=12.0cm,height=12.0cm}}
\caption{X-ray $\nu F_{\nu}$ lightcurves ($\nu=10^{18}$ Hz) for the
three types of bursts, starting from $t_{dec}$ and ending at the time
when the bulk Lorentz factor $\Gamma=2$. Curves I, II, and III are
calculated for regimes where the proton synchrotron, the electron IC,
and the electron synchrotron dominate, respectively.
All the parameters adopted are the same as those in Fig.2, except for
the regime II burst, where a slight denser medium $n=5 {\rm cm^{-3}}$
is adopted to show how the IC component flattens the light curve at
later times. For the regime II, the contributions from the three
different components are also shown with the same notations as in Fig.2.}
\end{figure*}

\begin{figure*}
\centerline{\psfig{file=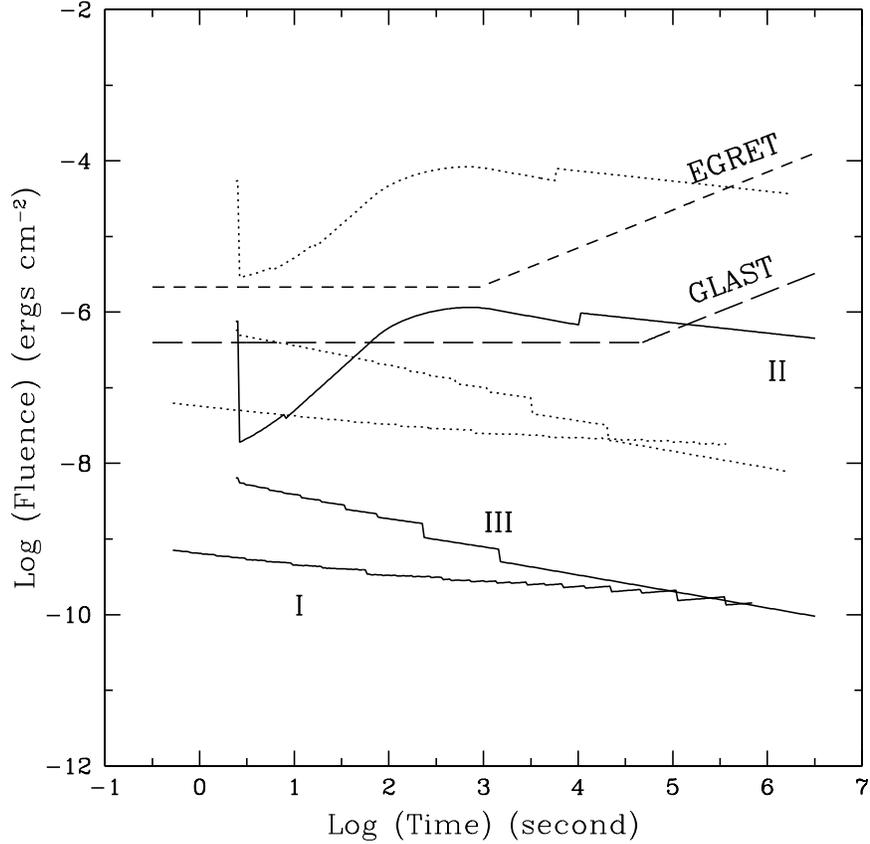,width=12.0cm,height=12.0cm}}
\caption{GeV $\int_{\nu_1}^{\nu_2}\nu F_{\nu}d\nu \cdot t$ lightcurves
for the three types of bursts, starting from $t_{dec}$ and ending at the
time when the bulk Lorentz factor $\Gamma=2$. The energy flux has been
integrated within the range of $\nu_1\sim 400$ MeV to $\nu_2\sim 200$ GeV
in order to compare with observations. The solid curves I, II and III
indicate bursts in regimes I, II, and III, respectively, at a typical
cosmological distance ($z=1$, for a flat $\Lambda=0$ universe).
The three dotted unmarked curves are the same types of bursts located
at $z=0.1$. The other parameters adopted are the same as those in Fig.2.
The sensitivity curves for {\em EGRET} and {\em GLAST} are also
plotted. For low integration time regime, at least 5 photon detection
is required. 
For the regime II parameters, the electron IC component gives an extended
duration GeV emission, easily detectable by {\em GLAST} at $z=1$
(solid line). For $z=0.1$ (dotted lines), {\em EGRET} can detect a
regime II burst, and this may account for the extended GeV detection
from GRB 940217.} 

\end{figure*}

\end{document}